\documentclass[times,twocolumn,tighten]{aastex63}
\usepackage{amsmath}
\usepackage{tablefootnote}

\newcommand{\mearth}{\,M$_\oplus$}

\newcommand{\msun}{\,M$_\odot$}

\definecolor{amber}{rgb}{1.0, 0.49, 0.0}
\definecolor{cadmiumorange}{rgb}{0.93, 0.53, 0.18}
\definecolor{chocolate}{rgb}{0.82, 0.41, 0.12}
\definecolor{oxfordblue}{rgb}{0.0, 0.13, 0.28}
\definecolor{firebrick1}{rgb}{0.698, 0.133, 0.133}

\graphicspath{{./}{figures/}}

\shorttitle{ALMA view of FU/EXor stars}
\shortauthors{Hales et al.}

\begin{document}

\title{\large ALMA Observations of Young Eruptive Stars: continuum disk sizes and molecular outflows}

\correspondingauthor{Antonio S. Hales}
\email{ahales@alma.cl}

\author[0000-0001-5073-2849]{Antonio S. Hales}
\affiliation{Joint ALMA Observatory, Avenida Alonso de C\'ordova 3107, Vitacura 7630355, Santiago, Chile}
\affiliation{National Radio Astronomy Observatory, 520 Edgemont Road, Charlottesville, VA 22903-2475, United States of America}
\author[0000-0003-2953-755X]{Sebasti\'an P\'erez}
\affiliation{Departamento de F\'isica, Universidad de Santiago de Chile, Av. Ecuador 3493, Estaci\'on Central, Santiago, Chile}
\author[0000-0003-4907-189X]{Camilo Gonzalez-Ruilova}
\affiliation{Departamento de Astronom\'ia, Universidad de Chile, Casilla 36-D, Santiago 8330015, Chile}
\affiliation{Nucleo de Astronom\'ia, Facultad de Ingenier\'ia y Ciencias, Universidad Diego Portales, Av. Ejercito 441, Santiago, Chile}
\author[0000-0002-2828-1153]{Lucas A. Cieza}
\affiliation{Nucleo de Astronom\'ia, Facultad de Ingenier\'ia y Ciencias, Universidad Diego Portales, Av. Ejercito 441, Santiago, Chile}
\author[0000-0001-5058-695X]{Jonathan P. Williams}
\affiliation{Institute for Astronomy, University of Hawaii, Honolulu, HI 96822, USA}
\author[0000-0002-9209-8708]{Patrick D. Sheehan}
\affiliation{National Radio Astronomy Observatory, 520 Edgemont Road, Charlottesville, VA 22903-2475, United States of America}
\author{Cristi\'an L\'opez}
\affiliation{Joint ALMA Observatory, Avenida Alonso de C\'ordova 3107, Vitacura 7630355, Santiago, Chile}
\author[0000-0002-0433-9840]{Simon Casassus}
\affiliation{Departamento de Astronom\'ia, Universidad de Chile, Casilla 36-D, Santiago 8330015, Chile}
\author[0000-0002-7939-377X]{David A. Principe}
\affiliation{MIT Kavli Institute of Astrophysics and Space Research, 70 Vassar St. Cambridge, MA, 02139, USA}
\author[0000-0002-5903-8316]{Alice Zurlo}
\affiliation{Escuela de Ingenier\'ia Industrial, Facultad de Ingenier\'ia y Ciencias, Universidad Diego Portales, Av. Ejercito 441, Santiago, Chile}
\affiliation{Nucleo de Astronom\'ia, Facultad de Ingenier\'ia y Ciencias, Universidad Diego Portales, Av. Ejercito 441, Santiago, Chile}

\begin{abstract}
We present Atacama Large Millimeter/submillimeter Array (ALMA) 1.3 mm observations of four young, eruptive star-disk systems at 0\farcs4 resolution: two FUors (V582~Aur and V900~Mon), one EXor (UZ~Tau~E) and one source with an ambiguous FU/EXor classification (GM~Cha). The disks around GM~Cha, V900~Mon and UZ~Tau~E are resolved. These observations increase the sample of FU/EXors observed at sub-arcsecond resolution by 15\%. The disk sizes and masses of FU/EXors objects observed by ALMA so far suggest that FUor disks are more massive than Class 0/I disks in Orion and Class~II disks in Lupus of similar size. EXor disks in contrast do not seem to be distinguishable from these two populations. We reach similar conclusions when comparing the FU/EXor sample to the Class~I and Class~II disks in Ophiuchus. FUor disks around binaries are host to more compact disks than those in single-star systems, similar to non-eruptive young disks. We detect a wide-angle outflow around GM~Cha in $^{12}$CO emission, wider than typical Class~I objects and more similar to those found around some FUor objects.  We use radiative transfer models to fit the continuum and line data of the well-studied disk around UZ~Tau~E. The line data is well described by a keplerian disk, with no evidence of outflow activity (similar to other EXors). The detection of wide-angle outflows in FUors and not in EXors support to the current picture in which FUors are more likely to represent an accretion burst in the protostellar phase (Class~I), while EXors are smaller accretion events in the protoplanetary (Class~II) phase.
\end{abstract}


\section{Introduction}\label{intro}

Multi-epoch accretion outbursts are believed to play a key role in the
build-up of the final stellar mass \citep{hartmann2008,
  hartmann2016}. Episodes of variable accretion are invoked to solve
the well-known luminosity problem, in which low-mass stars appear
fainter than predictions of steady-state accretion models
\citep{Kenyon1990,evans2009}. Despite being key to our understanding
of low-mass star (and planet) formation, the exact mechanisms that
trigger outbursts are still poorly understood, and this topic is of
growing interest as observational capabilities at long wavelengths
(such as ALMA) have developed \citep[see][for a review]{audard2014}.

Outbursting sources have been divided into two classes, FUors and
EXors (named after prototypes FU Ori and EX Lupi respectively).  FUors
have large ($\Delta$V$_{\rm mag}$ $\sim$ 5), long-lived (years to
decades) outbursts \citep{herbig1966}, whereas EXors have moderate
($\Delta$V$_{\rm mag}$ $\sim$ 2-4), shorter (days/months) episodes of
high accretion.  Several observational signatures indicate differences
in physical structures and accretion processes of each class, which
suggest that FUors and EXors might correspond to different
evolutionary stages.

EXors show rich optical and infrared emission lines similar to those
of classical T Tauri stars (CTTS), which brighten during the burst,
consistent with magnetospheric accretion. They also have spectral
energy distributions (SEDs) similar to those of Class II sources
\citep[e.g.][]{sipos2009}, suggesting these are typical star+disk
systems with no detectable remnant envelope. The few available optical
spectra of FUors in quiet states show they also resemble those of CTTS
\citep{Herbig1971}. In outburst, however, FUors show little signs of
magnetospheric accretion and instead exhibit optical P-cygni profiles
at H$\alpha$ and sodium lines \citep[indicative of strong winds of
  $\sim$10\% the accretion rate;][]{Calvet1993}, as well as double
peaked absorption features in some optical and near-IR lines. The
extremely high luminosity and spectroscopic features of the prototype
FU~Orionis itself have been explained in the frame of an accretion
disk model in which the disk is internally heated through strong
viscous accretion, overwhelming the stellar photosphere
\citep[][]{Hartmann1985,hartmann1996}.  This explains the high
accretion rates and double, differential, absorption lines \citep[with
  the difference in line profiles explained by absorption at different
  radii, e.g.][]{zhu2009a}. Most FUors are also associated with
reflection nebulae, and many have Class I SEDs with prominent CO
outflows observed at millimeter wavelengths, both indicating that they
are still partially surrounded by their parent envelope and therefore
suggesting a younger evolutionary stage compared to EXors
\citep[see][for review]{Reipurth2010,hartmann2016}.  Objects that
share the spectroscopic features of FUors, but for which outbursts
have never been observed are called FUor-like objects \citep[see ][for
  details]{connelley2018}.

Different mechanisms have been proposed to explain the origin for the
outbursts \citep{audard2014}: disk fragmentation plus subsequent
inward migration of clumps \citep{vorobyov2015}, a combination of
magnetorotational and gravitational instabilities
\citep{armitage2001}, and enhanced accretion induced by stellar
\citep{Bonnell1992} or planetary companions \citep{lodato2004}. The
key ingredients to distinguish between the above outburst mechanisms
are the total disk mass and the disk spatial structure. However, only
a few FU/EXor objects have been recently observed at sufficient
angular resolution at millimeter/submillimeter wavelengths. Recent
surveys with ALMA and the SMA \citep{cieza2018,liu2018} have shown
that 1.3~mm fluxes of the outbursting sources span over three orders
of magnitude, but the FUor objects are significantly brighter than the
EXor objects and typical Class II disks.

The inferred disk masses for the brightest FUor objects are large
enough that they may be gravitationally unstable.  Nevertheless,
follow up observations at 0\farcs03 (12 au) resolution of V883\,Ori,
the most massive disk in the survey, were unable to identify the
predicted signatures of instabilities or fragmentation (e.g. spirals
or clumps). The fainter targets are all EXor objects and have low disk
masses, $\sim$1-5~$M_{\rm Jup}$, that imply gravitational instability
is unlikely to play a role in their outbursts
\citep[e.g.,][]{cieza2018}.

Some FUor objects are close binaries with both components hosting
disks \citep[e.g. FU Orionis, L1551~IRS~, HBC~494;][]{hales2015,
  cruz2019, zurloprep}, whose interaction could help explain
the outbursts in some of the systems. Recent ALMA observations
  of FU Orionis show that the disks are indeed compact in continuum
  emission (11 au in radius) while the gas kinematics displays
  extended features possibly tracing binary and/or intra-cloud
  interactions \citep{perez2020}.

ALMA spectral line observations of $^{12}$CO show that FUors have
active circumstellar environments characterized by strong outflows
interacting with larger-scale structure \citep{ruiz2017a, ruiz2017b,
  zurlo2017, kospal2017b, principe2018, takami2019}.  On the other
hand, EXor sources do not show detectable outflows, with the possible
exception of V1647 Ori, a system with an unclear FUor/EXor
classification \citep{principe2018} and EX~Lupi itself \citep[around
  which a small arc-shaped feature is detected at $\sim$2 ~km~s$^{-1}$
  from the systemic velocity;][]{hales2018}. Although the number of
observed sources is small, the differences in outflow activity between
FUors and EXors suggest that the two types of objects represent an
evolutionary sequence comparable to normal Class~I and Class~II
respectively.  V346~Nor has observational properties more similar to
Class 0/I protostars, consistent with the growing number of Class 0
sources discovered to show eruptive behavior \citep{Safron2015,
  Johnstone2018}. This suggests that episodic accretion may play an
important role even at earlier protostellar stages. Studying the
larger-scale structure of eruptive sources is thus crucial for
understanding the nature of this common, yet short-lived, high mass
accretion variability in young stars.

This work aims to contribute a better understanding of the differences
between the two classes of young eruptive stars, FUors and EXors, and
how their disk masses, sizes and associated gas emission, compare to
Class I and Class II protostars. Are FU/EXor disks massive enough to
trigger gravitational instability? Do they show the predicted
signatures of a gravitationally unstable disk?  Are all variable
accretion sources associated with large molecular outflows? For this
purpose we present ALMA observations of four eruptive sources to
characterize their distribution of dust and gas, and to compare them
to those of non-eruptive protostars.  Section~\ref{obs} describes our
target sample, observations and data reduction. In
Section~\ref{results} we present our results, which are discussed in
Section~\ref{discussion}.  Section~\ref{conclusion} presents our
conclusions.

\begin{deluxetable*}{lcccccccccc}
\tablecaption{Summary of ALMA Observations (this work) \label{log2}}
\tablewidth{700pt}
\tabletypesize{\scriptsize}
\tablehead{
\colhead{Name} & 
\colhead{Execution Block} & 
\colhead{N Ant.} & 
\colhead{Date} & 
\colhead{ToS } & 
\colhead{Avg. Elev. } & 
\colhead{Mean PWV } & 
\colhead{Phase RMS } & 
\colhead{Baseline } & 
\colhead{AR } & 
\colhead{MRS} \\
\colhead{} & 
\colhead{} & 
\colhead{} & 
\colhead{} & 
\colhead{(sec)} & 
\colhead{(deg)} & 
\colhead{(mm)} & 
\colhead{(rad)} & 
\colhead{(m)} & 
\colhead{(")} & 
\colhead{(")} 
}
\startdata
V582 Aur &uid://A002/Xd248b5/X4ff9&41&2018-09-22 &989&30.3 &0.2&0.435&15.1-1397.8 &0.3&3.6\\
V900 Mon&uid://A002/Xd23397/Xe02d &43&2018-09-21 &1128&75.4&0.2&0.411&15.1-1397.8 &0.3&3.7\\
UZ Tau   &uid://A002/Xd23397/X41f6&47&2018-09-20 &1273&31.2&0.4&0.472&15.1-1397.8 &0.3&3.8 \\
GM Cha&uid://A002/Xd28a9e/X4a7b   &43&2018-09-27 &1417&36.4&1.5&0.498&15.1-1397.8 &0.3&3.6 \\
GM Cha&uid://A002/Xd29c1f/X5b5e   &43&2018-09-29 &1445&36.3&1.0&0.458&15.1-1397.8 &0.2&3.3 \\
%B3    &uid://A002/Xc4b006/X1b88   &42&2017-09-21 &2453&73.7&1.6&0.164&41.4-12145.2&0.1&1.2 \\
%B7    &uid://A002/Xc44eb5/X7991   &47&2017-09-08 &3309&71.7&0.5&1.222&21.0-10636.0&0.0&0.8
\enddata \tablecomments{Summary of the new ALMA observations
  presented in this work, including number of antennas, total time on
  source (ToS), target average elevation, mean precipitable water
  vapor column (PWV) in the atmosphere, phase RMS measured on the
  bandpass calibrator, minimum and maximum baseline lengths, expected
  angular resolution (AR) and maximum recoverable scale (MRS).}
\end{deluxetable*}

\begin{deluxetable*}{lcccc}
  \tablecaption{Summary of continuum disk detections. Disks are sorted
    by declining disk mass. Object type, Spectral type, and L$_{\rm
      Bol}$, and Companions are taken from
    \citet{audard2014}. Distances are obtained from the second data
    release (DR2) of Gaia \cite[]{gaia2018}, except for GM~Cha for
    which we adopt the distance to Cha~I from
    \citet{Dzib2018}. \label{tableobs}}
    
\tablewidth{700pt}
\tabletypesize{\scriptsize}
\tablehead{
\colhead{} & 
\colhead{V582 Aur} & \colhead{V900 Mon} & 
%\colhead{GM Cha X4a7b} &
%\colhead{GM Cha X5b5e} &
\colhead{UZ Tau E} & 
\colhead{GM Cha}
}
\startdata
Object type  & FUor & FUor & EXor & FU/EXor \\
%Spectral type & - & - & - & - & - & - \\
L$_{\rm Bol}$(L$_{\odot}$)  & - & 106 & 1.7 & $>$1.5  \\
Companions & - & N & Y (SB+4$''$) & Y (10$''$)  \\
F$_{\rm 1.3mm}$ (mJy)\tablenotemark{a}  & 5.3$\pm$0.6 & 9.8$\pm$0.1 & 134$\pm$1 & 10.4$\pm$0.1\\
Major axis (mas)\tablenotemark{b} & - & 72$\pm$11  & 668$\pm$20 & 613$\pm$8\\
Minor axis (mas)\tablenotemark{b} & - & 60$\pm$20 & 396$\pm$22 & 221$\pm$2\\
Position angle (deg) & - & 164$\pm$63 & 90$\pm$4 & 27.8$\pm$0.3\\
Disk radius (au) & - & 54 & 44 & 49\\
Inclination (deg) & - & 50 & 59 & 70\\
Distance (pc) & 2575 & 1500 & 131 & 192 \\
M$_{\rm dust}$ (M$_{\oplus}$) for $T=20~K$ & 1055 & 662 & 69 & 11 \\
M$_{\rm dust}$ (M$_{\oplus}$) for $T=60~K$ & 291 & 182 & 19 & 3 \\
M$_{\rm disk}$ (M$_{\rm Jup}$) for $T=20~K$ & 332 & 208 & 22 & 3 \\
Peak (mJy~beam$^{-1}$) & 4.2 & 9.4 & 54.7 & 5.9 \\
Rms (mJy~beam$^{-1}$) & 1.41$\times$10$^{-1}$ & 5.23$\times$10$^{-2}$ & 8.08$\times$10$^{-2}$ & 3.56$\times$10$^{-2}$\\
\hline
Beam Major axis (") &  0.49 & 0.35  & 0.57  &0.57 \\
Beam Minor axis (") &  0.28 & 0.32  & 0.34 &0.37  \\
Beam Position angle (deg) & 153.2 & 77.5  & -40.1 &17.2 \\
\enddata
\tablenotetext{a}{Uncertainties on the continuum flux do not include the absolute flux uncertainty of ALMA.}
\tablenotetext{b}{Deconvolved disk sizes reported by {\sc IMFIT}.}

\end{deluxetable*}

\section{Sample Selection and ALMA Observations}\label{obs}

\subsection{Target Sample}\label{sample}

There are about $\sim$38 known FUor/EXor objects within $\sim1$~kpc,
of which $\sim$20 are observable with ALMA \citep[with declination
  $<+40^\circ$;][]{audard2014}. So far $\sim$14 of them
have already been observed by ALMA at moderate angular resolution
\citep[0\farcs2-0\farcs9;][]{hales2015, kospal2017b, cieza2018,
  hales2018, takami2019, cruz2019}. In this work we present ALMA Band
6 observations of four FUor/EXor objects observed in ALMA Cycle 5
(project code 2017.1.01031.S, PI Hales). The targets were selected
from the list of \citet{audard2014} and include two FUors (V582~Aur
and V900~Mon), one EXor (UZ~Tau~E) and one source with ambiguous
FU/EXor classification (GM~Cha).  This apparently small
  sample represents 10\% of known FUor/EXor objects, and these
  observations increase the number of eruptive sources observed at
  sub-arcsecond resolution by 15\% \citep[since two of our targets,
    V900~Mon and UZ~Tau~E, have already been observed at similar or
    higher angular resolution;][]{takami2019, long2018}.

V900~Mon is an FUor object discovered by \citet{Thommes2011},
who reported a brightening of at least 4 magnitudes in optical
magnitude, while follow up observations indicate the object
shows FUor characteristics such as P-Cygni profiles in H$\alpha$ and
sodium lines, CO absorption in near-IR, as well as an association with
a bright, compact reflection nebula
\citep{Reipurth2012}. \citet{takami2019} presented recent ALMA
observations that show the presence of a CO outflow and rotating
envelope.

V582~Aur is young eruptive system first identified due to an optical
brightening of $\sim$4 magnitudes that took place sometime between
1982 and 1986, followed by the appearance of a nebula that was not
visible in previous optical images
\citep{Samus2009}. \citet{Semkov2013} studied the optical photometry
and spectra and showed that the star has spectroscopic signatures
typical of FUors, although some of the color changes have similarities
with UXor-type stars \citep[UXors are young stars that show stochastic
  variability than can be explained by eclipses due to dust fragments
  in their circumstellar disks;][]{Grinin2019}. \citet{abraham2018}
studied the color variations of V582~Aur and suggest that the two 1~yr
long dips seen in 2012 and 2017 in the light curve are due to variable
extinction rather than enhanced accretion, and therefore more similar
to UXors than to FUor/EXors. \citet{Zsidi2019} combine optical, near-
and mid-infrared photometry to investigate the physical structure of
the dust responsible for the dimming. Millimeter observations with
IRAM 30m and Northern Extended Millimeter Array (NOEMA) reveal a
compact continuum source at the position of V582~Aur, together with
other clumps in both continuum and CO isotopologues
\citep{abraham2018}.

GM~Cha (ISO-Cha I 192) is a Class I/II source located in the Chameleon
I dark cloud \citep{Moody2017,Mottram2017}. The source has no optical
counterpart, is located in a region of high extinction
\citep{jones1985}, and is associated with a $^{12}$CO outflow
\citep{mattila1989,Mottram2017}. From 1996 to 1999 the K$_{S}$-band
magnitude increased by 2 magnitudes \citep{persi2007}, and there is
evidence of possible elongated infrared nebulosity in the direction of
the outflow. As with V1647~Ori, the observational characteristics of
GM~Cha are similar to both FUors and EXors. The SED can be well
described by a star+disk+infalling envelope and the presence of a
reflection nebula suggests similarities with FUors. However, it shows
no 2.3~$\mu$m CO band head (neither in absorption or emission), and
the derived accretion rate of 10$^{-7}$~M$_{\odot}$~yr$^{-1}$ is three
orders of magnitude lower than those of FUors \citep{persi2007}.

UZ~Tau is a well-studied quadruple system, containing a spectroscopic
binary with a separation of $\sim$0.03 au \citep{mathieu1996}
surrounded by a large circumbinary disk \citep[UZ Tau
  E;][]{tripathi2018}, and another M3+M3 binary pair with a 0\farcs34
($\sim$48 au) projected separation (UZ Tau W). UZ~Tau~E continuum disk
was imaged by \citet{long2018} at 0\farcs13$\times$0\farcs11
resolution and found the presence of at least three sets of rings
spanning from 0 to $\sim$100~au with one clearly defined gap at
69~au. UZ~Tau~E shows moderate ($\sim$1~mag) short-term variability in
the optical and infrared, and exhibits characteristics of an EXor
\citep{Lorenzetti2007}. \citet{Jensen2007} showed that the periodic
variability of UZ~Tau~E could be explained by variable accretion
caused by interactions between the binary orbit and the circumbinary
disk. \citet{Czekala2019} used ALMA data of $^{13}$CO and C$^{18}$O to
study the degree of alignment of the binary and the circumbinary disk,
and determined that the disk and the stars are nearly coplanar, which
may imply that planets formed in this system will also be coplanar.

\subsection{Observations}

ALMA observations of these four targets were acquired between
September 20$^{th}$ and September 29$^{th}$ 2018 using the Band~6
receiver ($\sim$~230~GHz). The total number of available 12 meter
antennas ranged from 41 to 47, providing baselines ranging from 15.1~m
to 1.397~km.  A summary of the observations such as precipitable water
vapor column (PWV) in the atmosphere, phase RMS, target elevation and
time on source (ToS), number of antennas, expected angular resolution
(AR) and maximum recoverable scale (MRS) are presented in
Table~\ref{log2}. Standard observations of bandpass, flux and phase
calibrators were also included.

The spectral setup was chosen to target the $^{12}$CO(2--1),
$^{13}$CO(2--1) and C$^{18}$O(2--1) transitions of carbon monoxide
(rest frequencies of 230.538~GHz, 220.399~GHz, 219.560~GHz, and
218.437~GHz, respectively). The ALMA correlator was configured in
Frequency Division Mode (FDM) to provide spectral resolutions of
0.09~km~s$^{-1}$. Two spectral windows in Time Division Mode (TDM) to
image dust continuum were set up at central frequencies of 218.0~GHz
and 233.0~GHz, each with total bandwidths of 1.875~GHz.

\subsection{Data Reduction}

All data were calibrated using the ALMA Science Pipeline (version
40896 Pipeline-CASA51-P2-B) in CASA 5.1
\citep[CASA{\footnote{\url{http://casa.nrao.edu/}}};][]{2007ASPC..376..127M}
by staff at the North American ALMA Science Center. The Pipeline uses
CASA tasks to perform the data reduction and calibration in a standard
fashion which includes correction for Water Vapor Radiometer (WVR) and
system temperature, as well as bandpass, phase, and amplitude
calibrations.

Imaging of the continuum and molecular emission lines was performed
using the {\sc tclean} task in CASA. The two continuum spectral windows
were imaged together using Briggs weighting with a robust parameter of
0.5, resulting in a final continuum image centered at 225.5~GHz. The
synthesized beam size achieved for each target are shown in
Table~\ref{tableobs}. Since all targets are bright enough for
self-calibration, a single iteration of phase-only self-calibration
was performed to improve coherence.  The resulting continuum sensitivity
achieved for each target is shown in Table~\ref{tableobs}.

Imaging of the spectral lines was performed using {\sc tclean} on the
continuum subtracted data (which subtracted using CASA task {\sc
  uvcontsub}). Self-calibration tables from the continuum data were
applied to the spectral line data before imaging the CO lines.  During
{\sc tclean}ing the spectral channels were binned to 0.3~km~s$^{-1}$
for UZ~Tau~E and to 0.5~km~s$^{-1}$ for the other sources.  {\sc
  tclean} was run with natural weighting to enhance sensitivity.

\section{Results}\label{results}

\begin{figure*}
  \centering
  \includegraphics[height=0.26\textwidth]{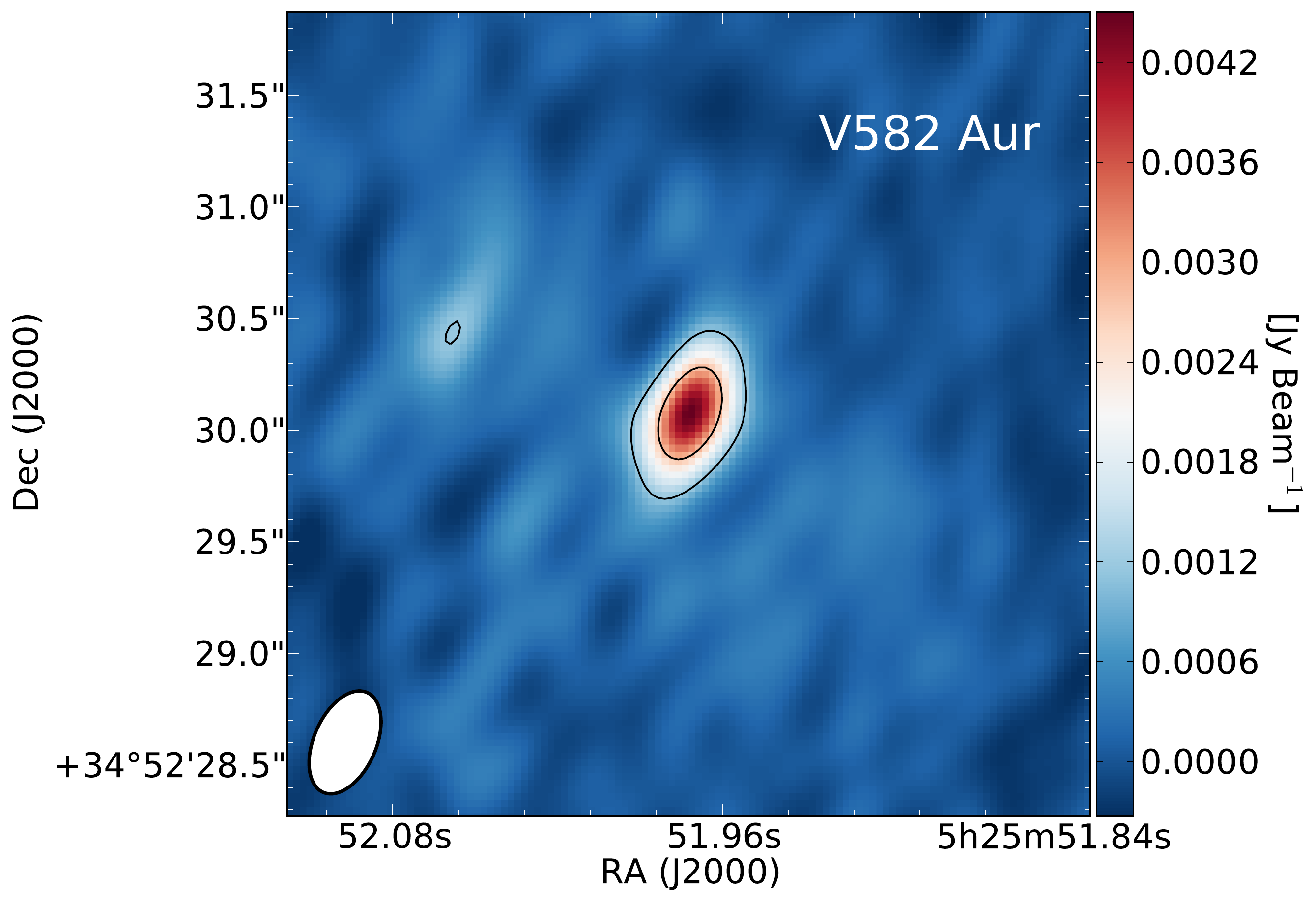}
  \includegraphics[angle=0,scale=0.33]{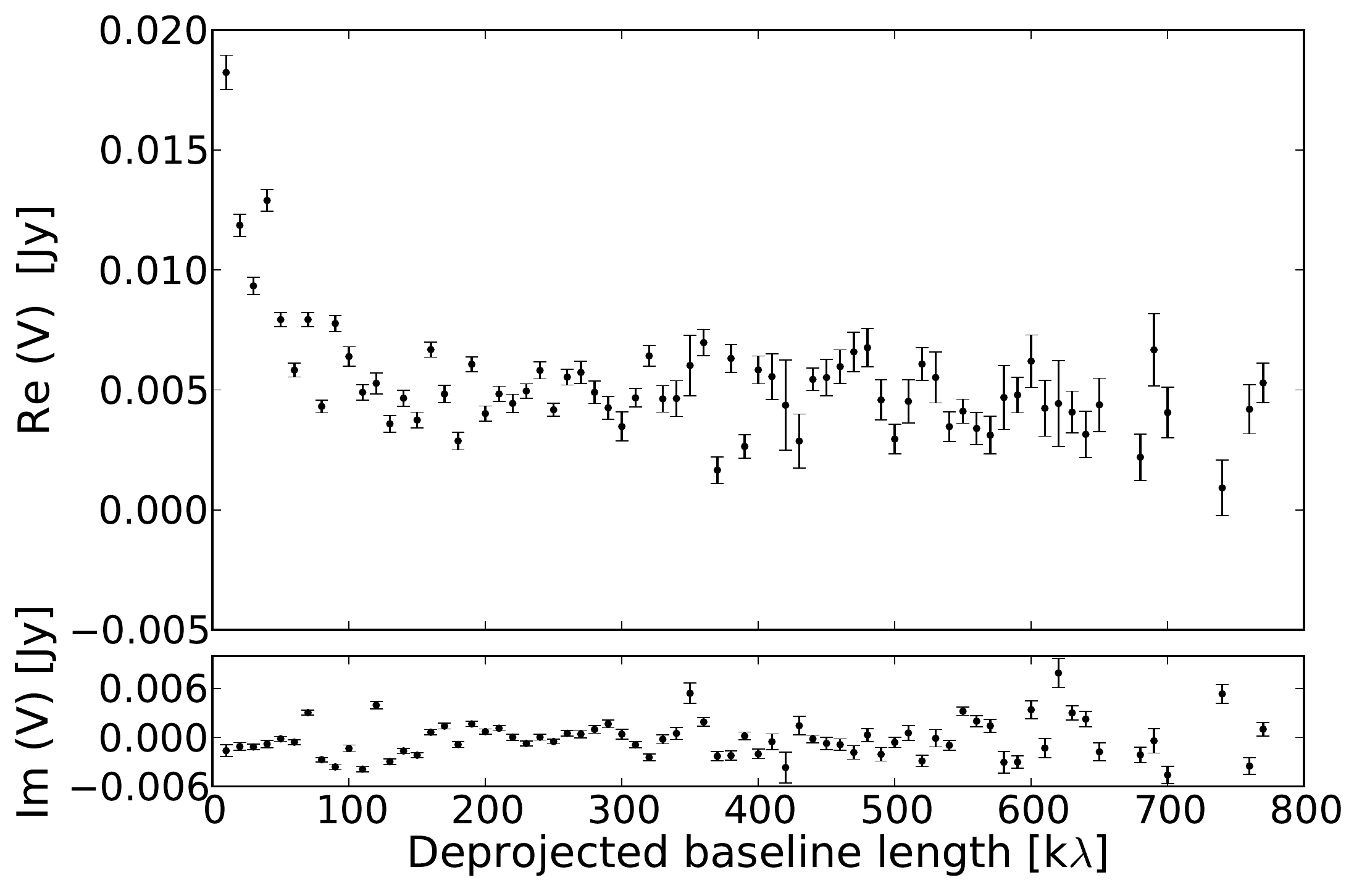}\\[0.2cm]
  \includegraphics[height=0.26\textwidth]{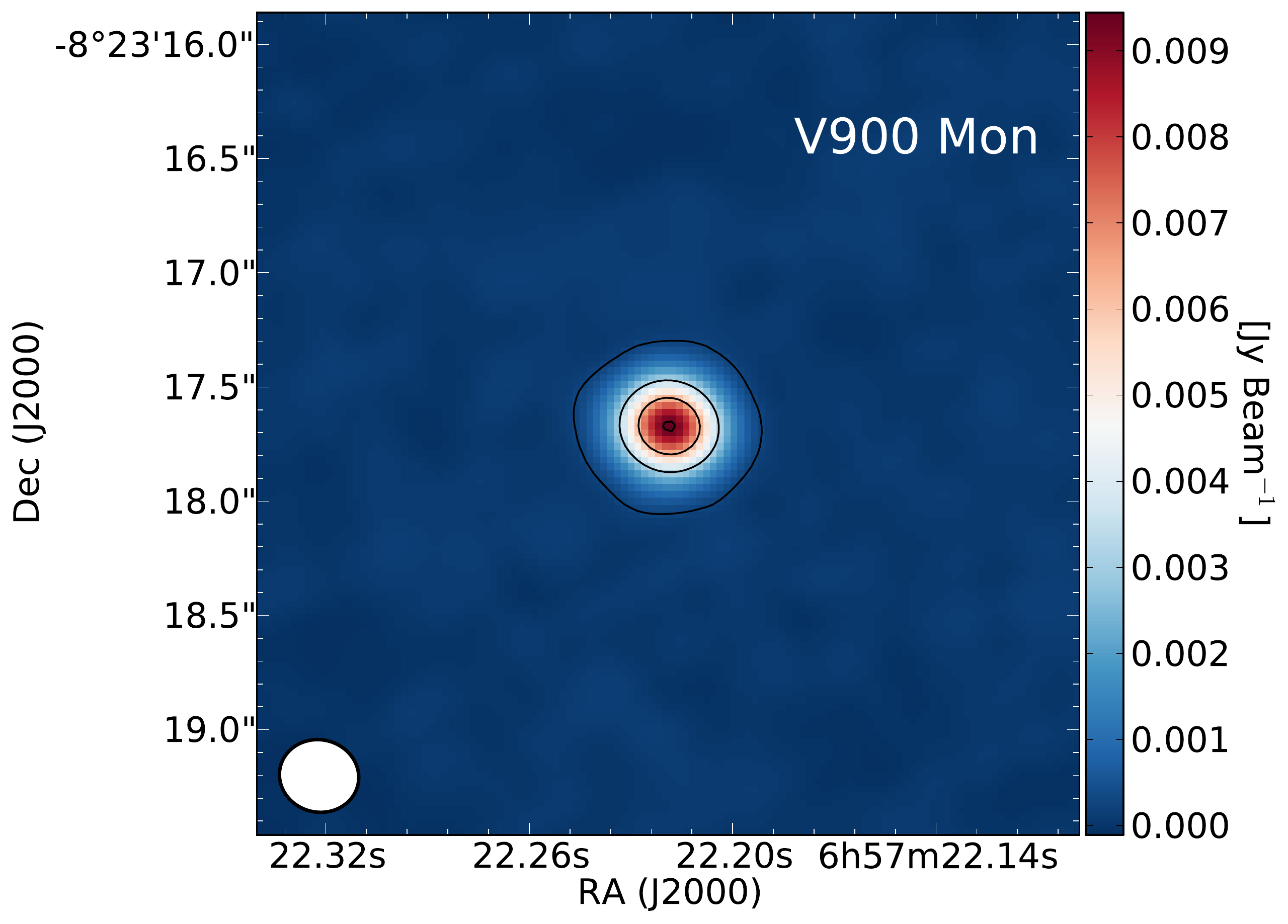}
  \includegraphics[angle=0,scale=0.33]{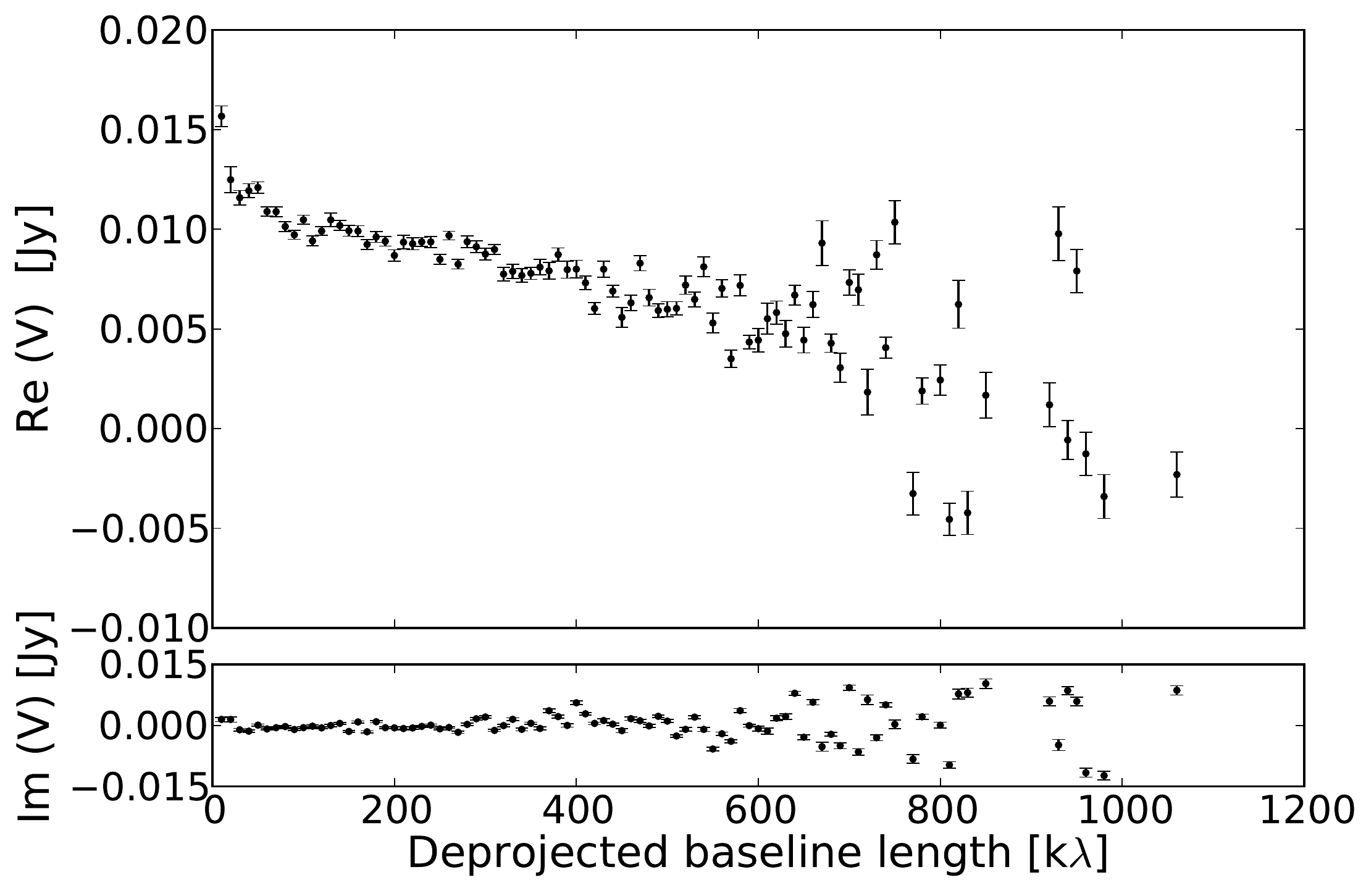}\\[0.2cm]
  \hspace*{-0.3cm}\includegraphics[height=0.26\textwidth]{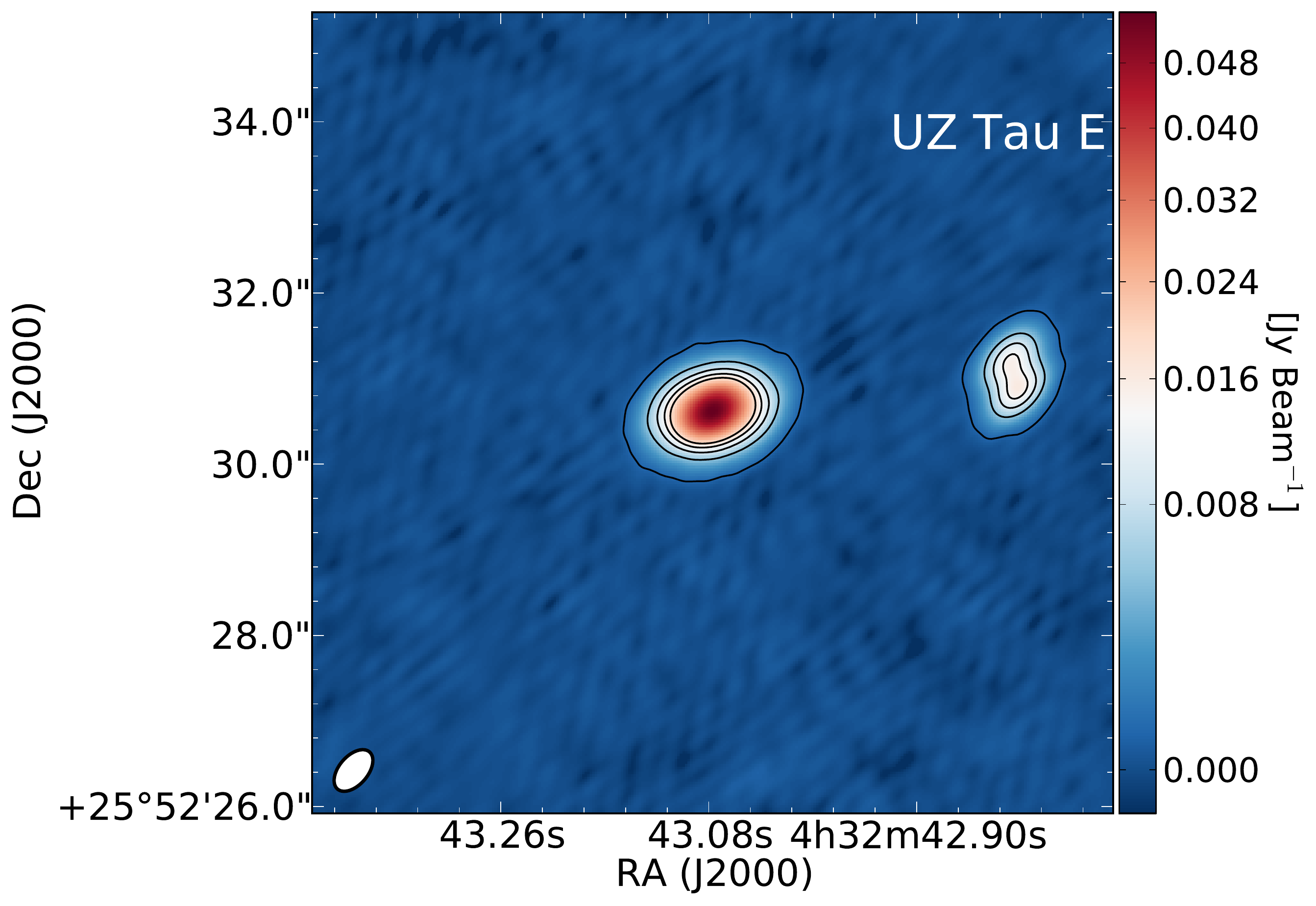}
  \includegraphics[angle=0,scale=0.33]{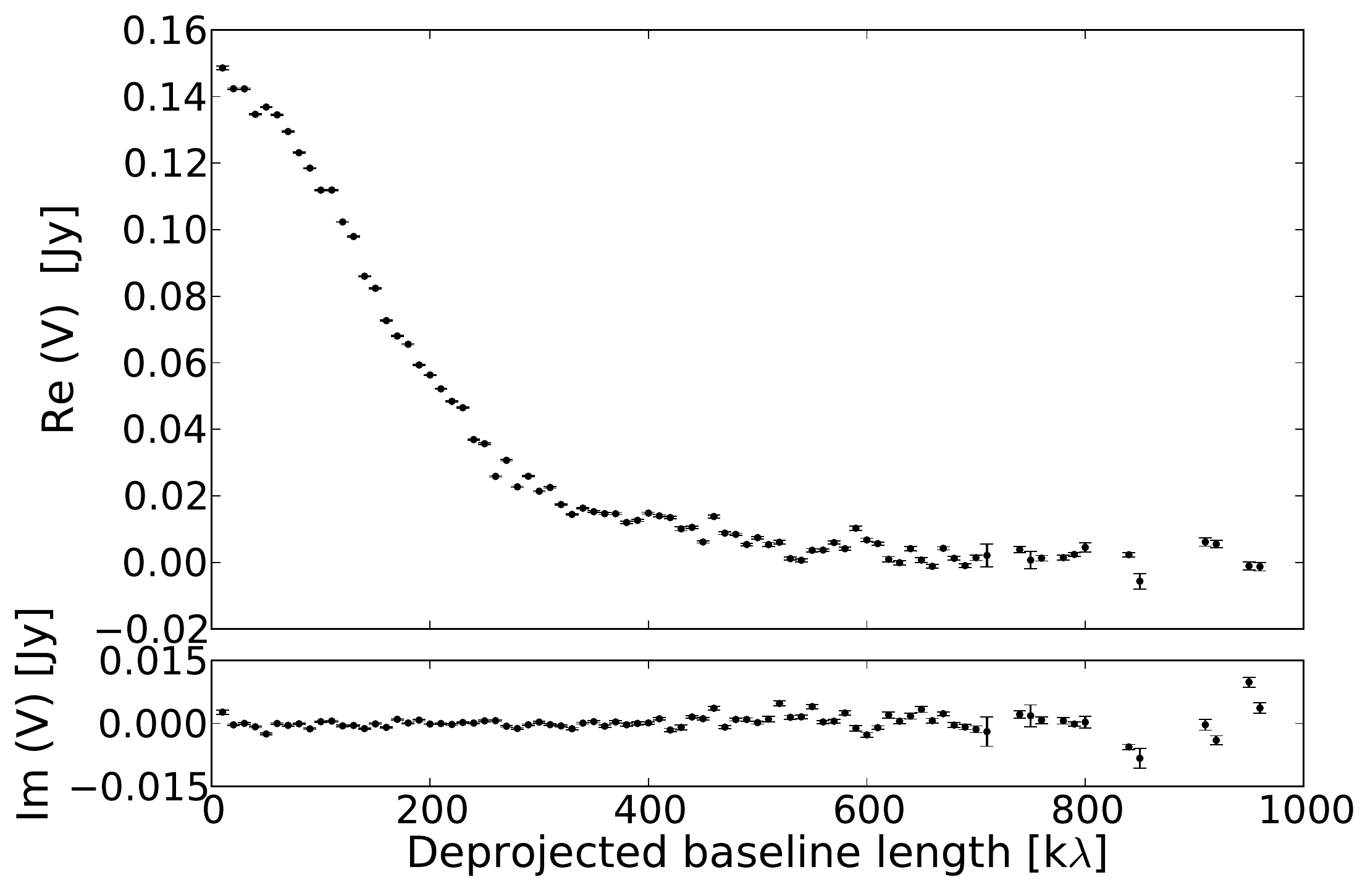}\\[0.2cm]
  \includegraphics[height=0.26\textwidth]{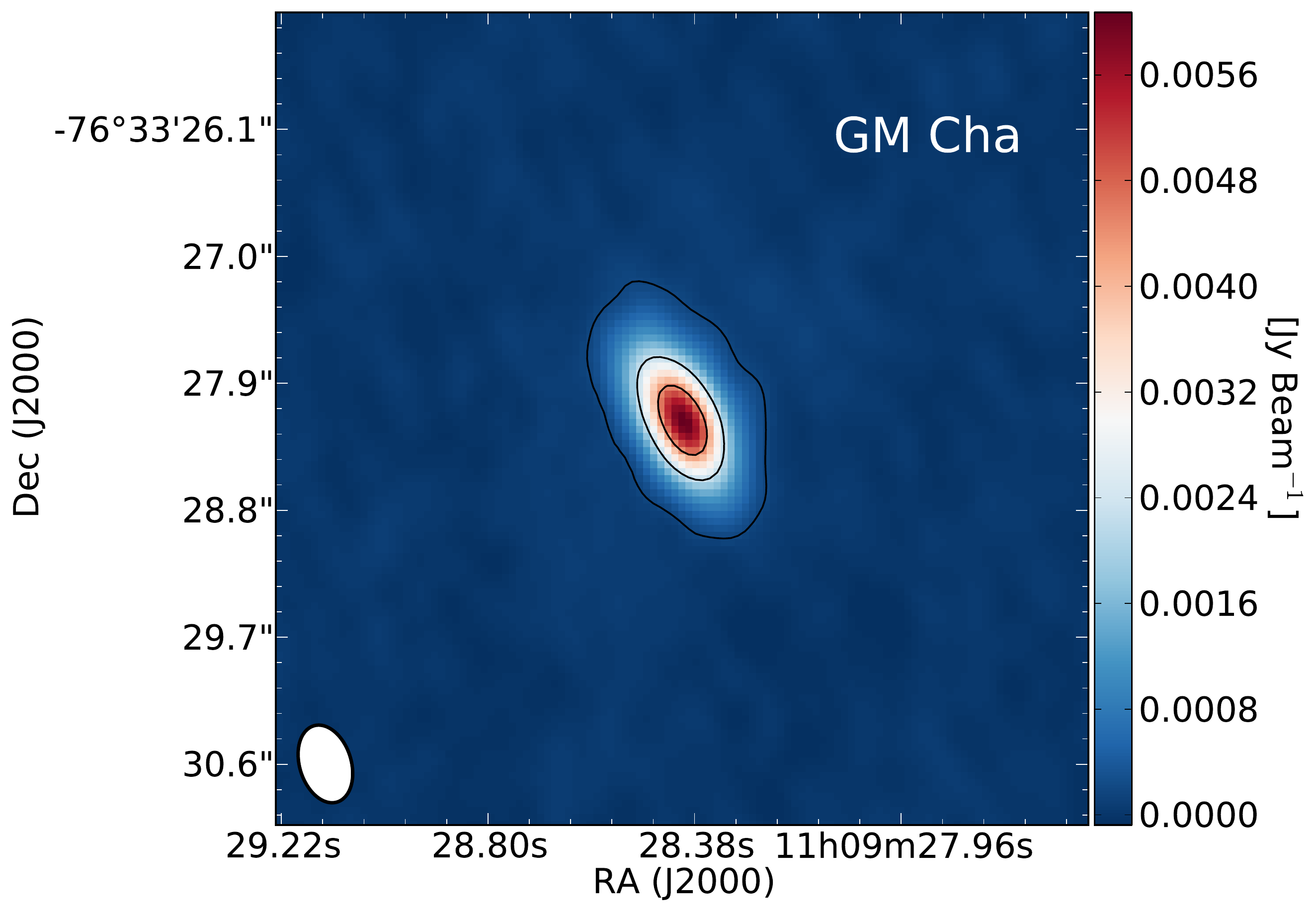}
  \includegraphics[angle=0,scale=0.33]{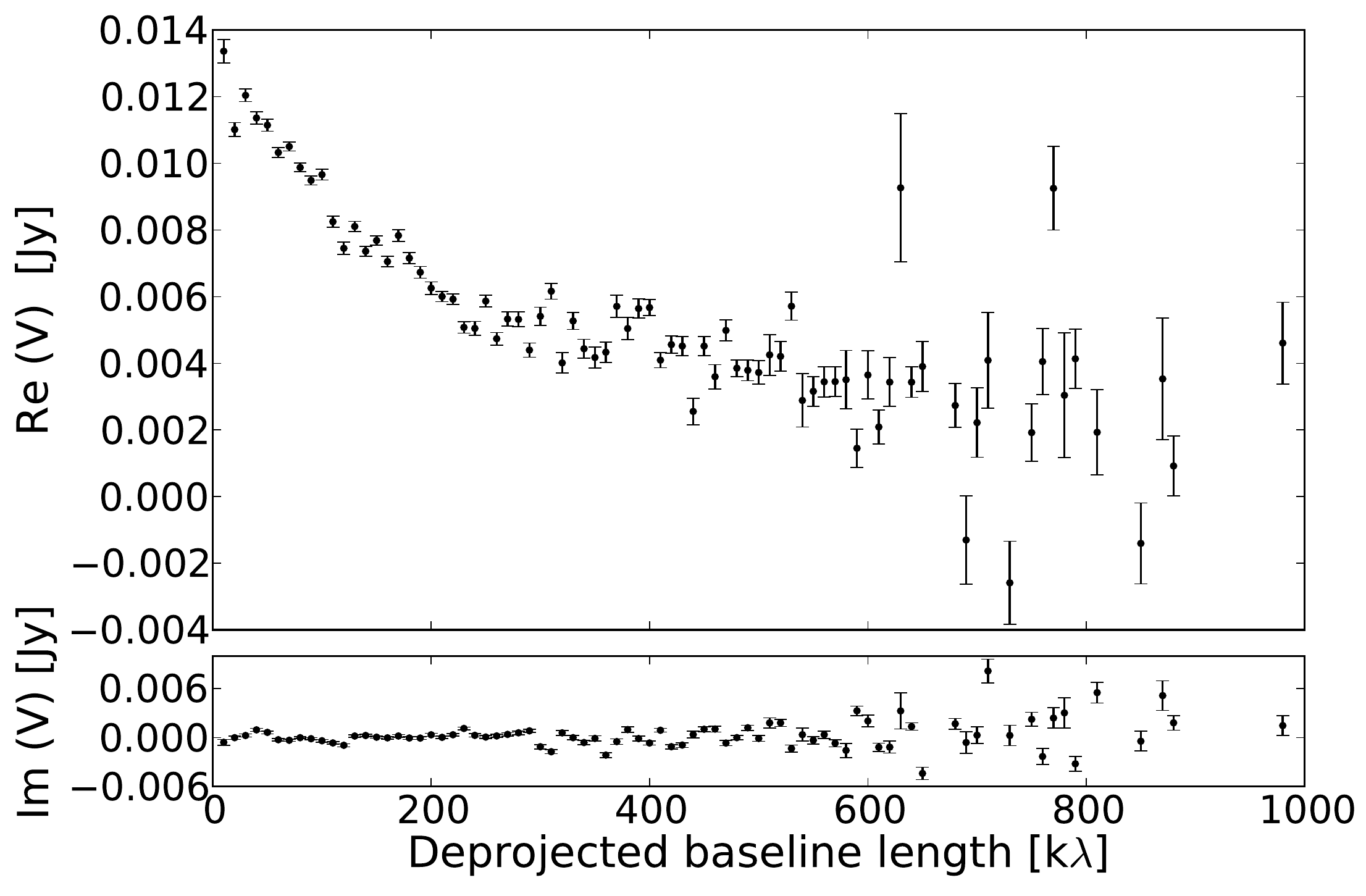}             

  \caption{ {\it{Left}}: ALMA 1.3mm continuum images for the FUor
    (top) and EXor (bottom) sources.  Contour levels for V900~Mon,
    UZ~Tau and GM~Gha start at 5$\sigma$ and increase in steps of
    5$\sigma$, and in steps of 10$\sigma$ for V582~Aur. The
    synthesized beam size achieved for each target are shown in the
    lower left.  {\it{Right}}:  Averaged continuum
    visibilities.  }
\label{continuumimages}

\end{figure*}

\subsection{Continuum}\label{contobs}

All observed targets are detected at a high signal-to-noise ratio in
continuum emission (Figure~\ref{continuumimages}).  We use the CASA
task {\sc imfit} to fit an elliptical Gaussian to the images and
derive emitting region sizes (deconvolved from the beam) and dust
continuum fluxes. We use these sizes as a proxy for the disk
radius. The disk sizes are thus estimated from the deconvolved
Gaussian FWHM/2 (see Section~\ref{masses} for a discussion on the
reliability of this method for estimating disk sizes). The disks
around GM~Cha and UZ~Tau~E are resolved. V900~Mon is marginally
resolved, as can be seen by inspecting the visibilities (shown in the
right panel of Figure~\ref{continuumimages}). V582~Aur, being
significantly more distant, is unresolved.  The large circumbinary
disk around UZ~Tau~E is resolved, while the individual disks around
each component of UZ Tau W are marginally resolved.  The derived disk
parameters are presented in Table~\ref{tableobs}.  The inclinations
are derived from the ratio of the deconvolved minor and major axes.
The 225.5~GHz flux for UZ Tau E is consistent with the CARMA
measurement of $131\pm 6$\,mJy \citep[][]{tripathi2018}, as well as
with 0.12'' resolution ALMA observations \citep[$129.5\pm
  0.2$\,mJy;][]{long2018}. The disk radius we derive for V900~Mon
(54~au) is consistent with the 45~au value inferred by
\citet{takami2019} with a factor $\sim$2 better spatial resolution.

The 1.3 mm photometry of each disk is listed in
  Table~\ref{tableobs}. These fluxes will be used to estimate dust
  masses using the optically thin approximation in
  Section~\ref{masses} along with a discussion of the key assumptions
  and their caveats. 

\subsection{Spectral line data }\label{linedata}

\begin{figure*}
  \includegraphics[angle=0,height=0.35\textwidth]{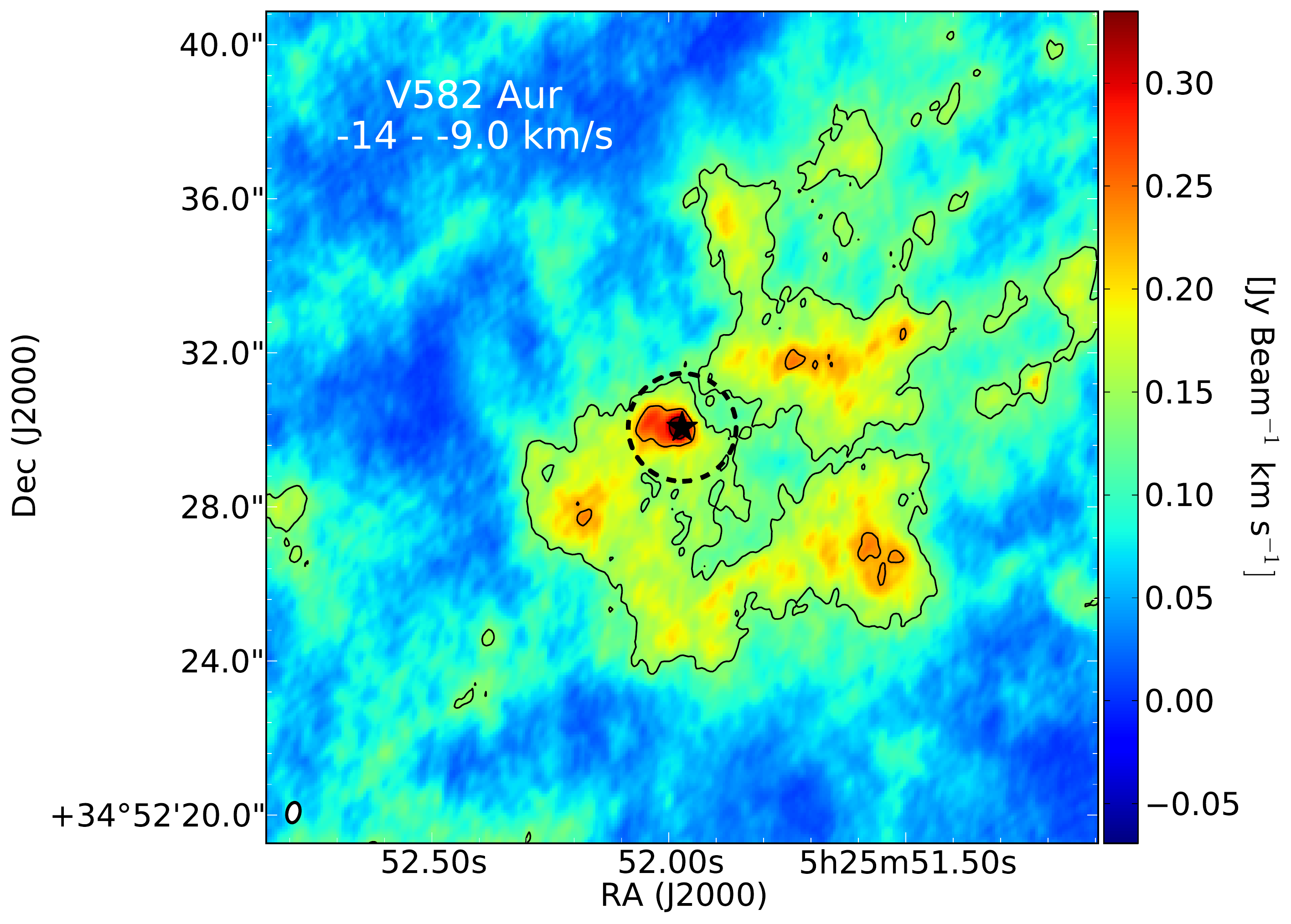}\hfill
  \includegraphics[angle=0,height=0.35\textwidth]{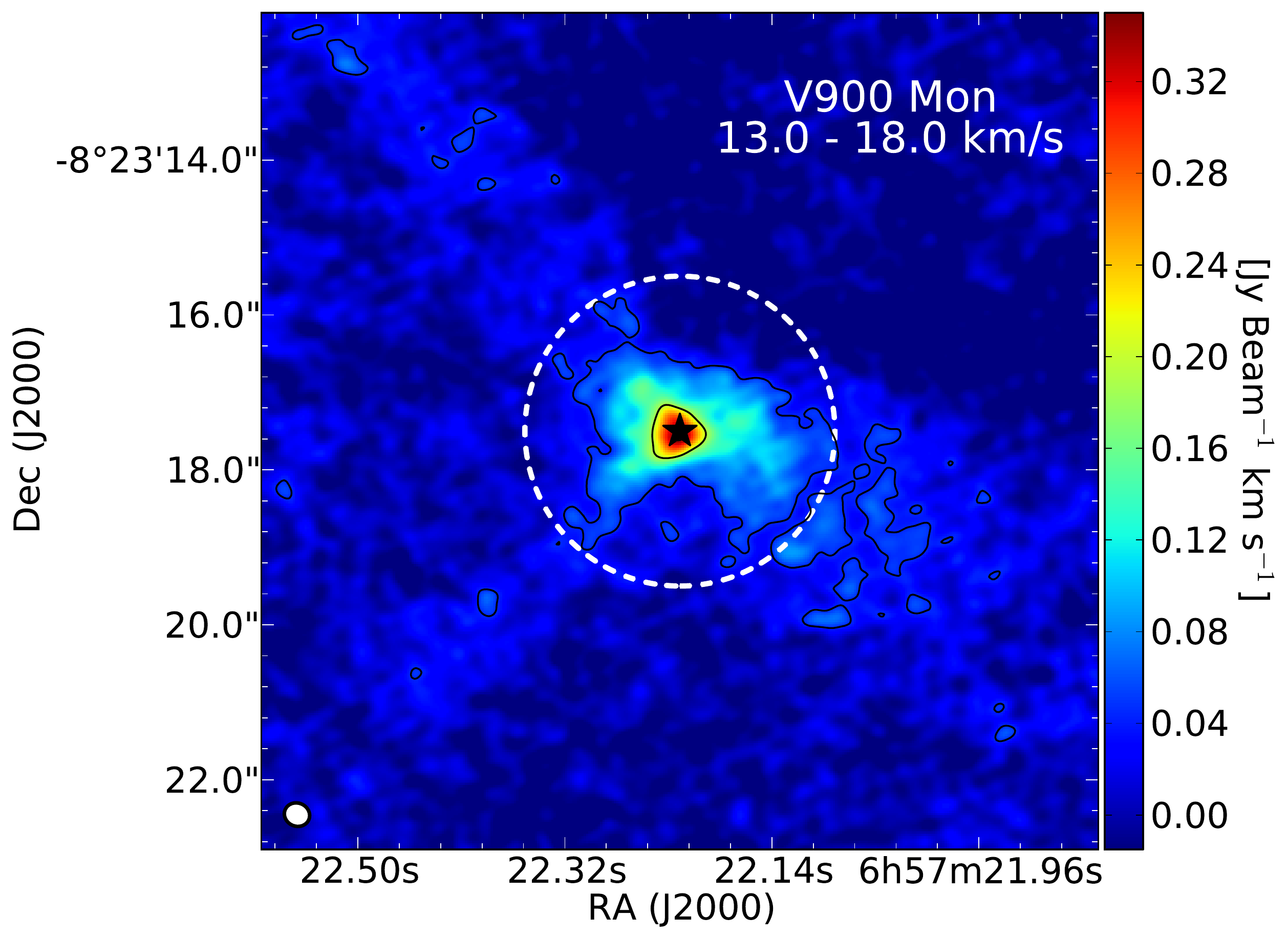}\\
  \includegraphics[angle=0,height=0.355\textwidth]{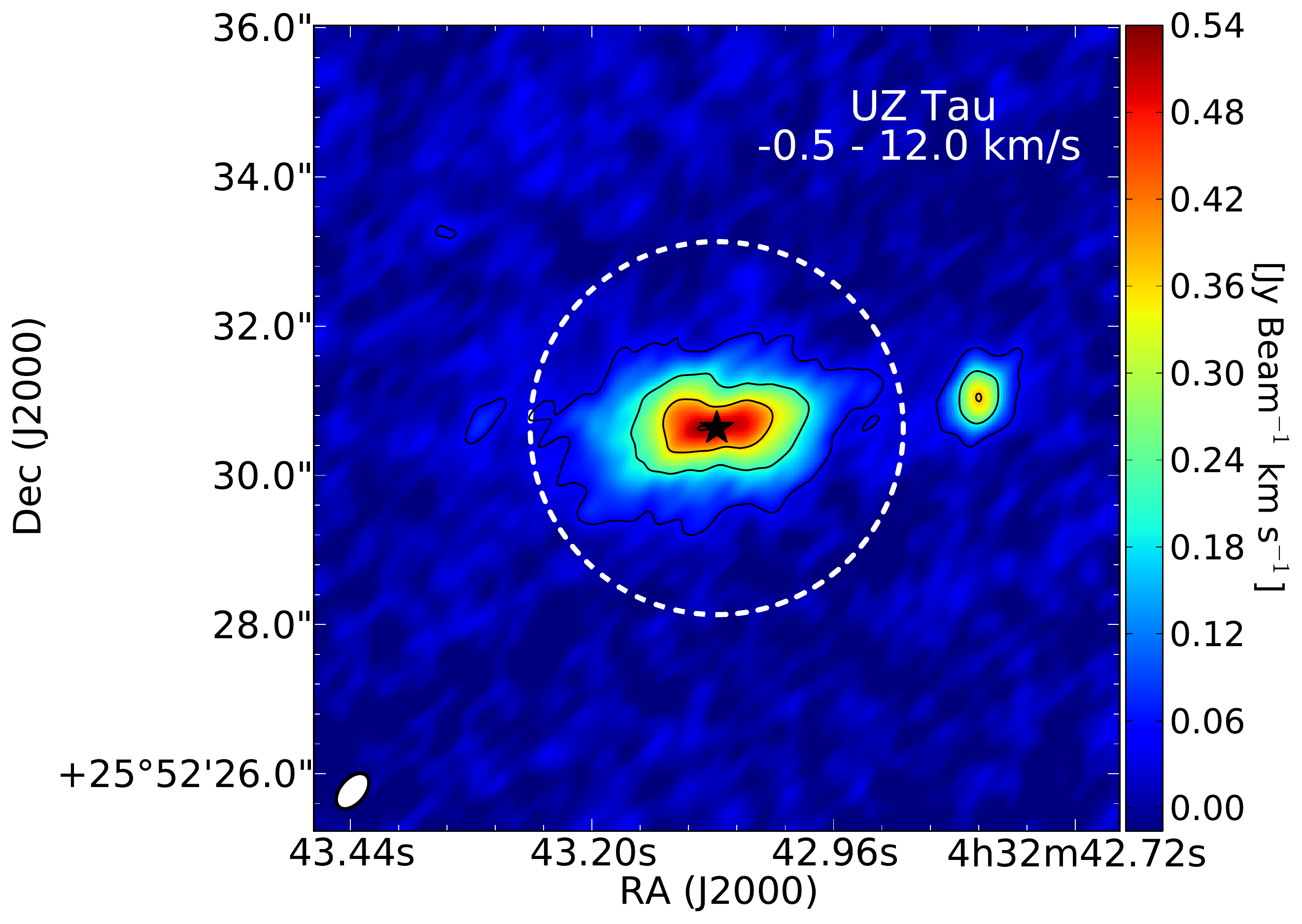}\hfill
  \includegraphics[angle=0,height=0.35\textwidth]{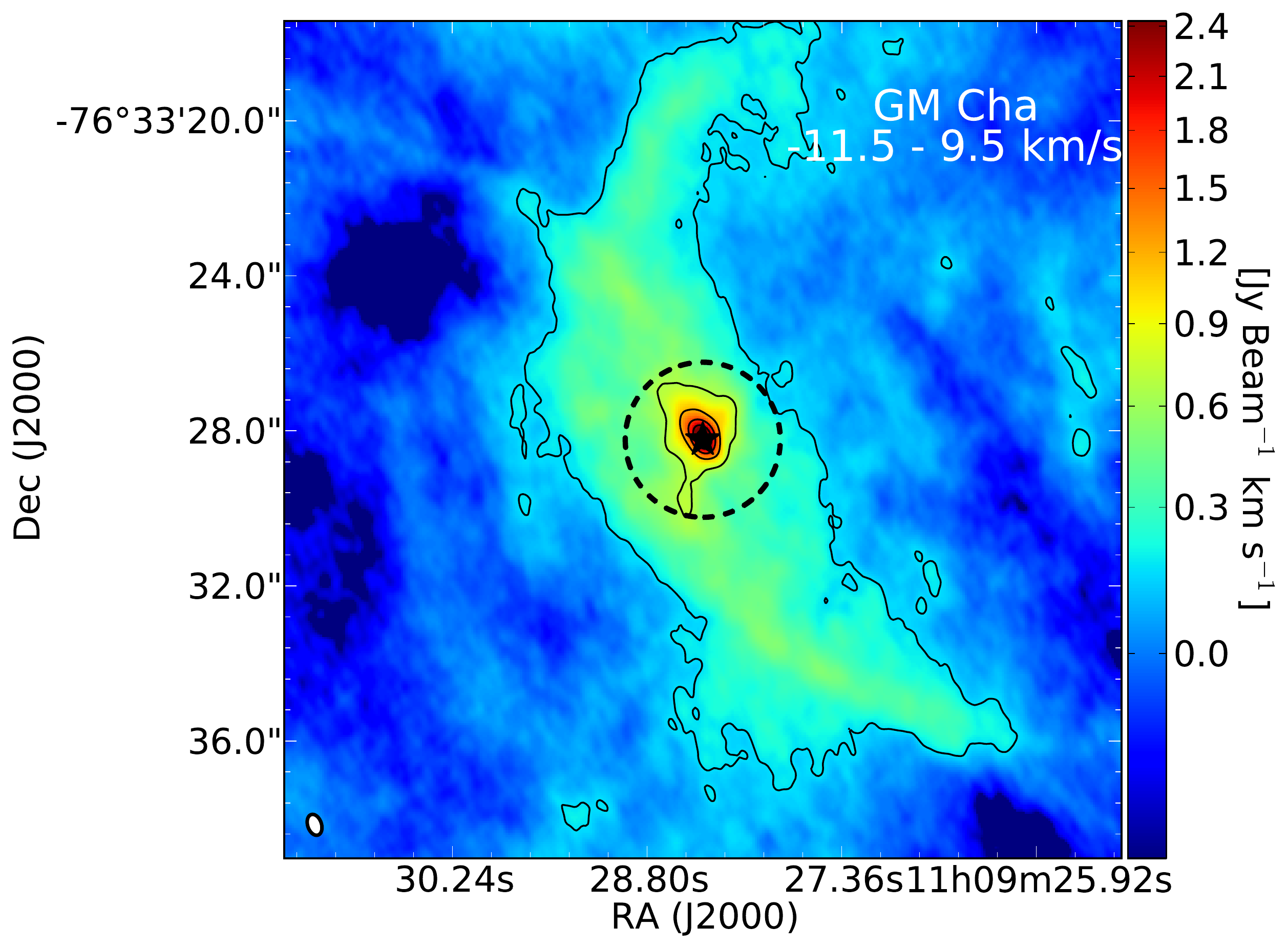}
  
  \caption{$^{12}$CO integrated intensity maps (moment 0) for all
    targets. The dashed circles shows the region used to compute the
    integrated line emissions listed in Table~\ref{cofluxes}. The peak
    position of the continuum is shown with a star.}
  \label{mom0}
\end{figure*}

Spectral line emission ($J$ = 2--1) from the three main isotopologues
of CO was detected in all sources. \citet{takami2019} recently
presented a study of V900~Mon in similar spectral transitions, and a
detailed analysis of the spectral line data for GM~Cha will be
presented in a separate paper \citep{gonzalesprep}. Therefore in this
work we present the $^{12}$CO data for all four sources for
comparison purposes, and the complete line data ($^{12}$CO, $^{13}$CO
and C$^{18}$O) only for UZ~Tau~E and V582~Aur.

\begin{figure*}
  \centering
  \includegraphics[angle=0,scale=0.3]{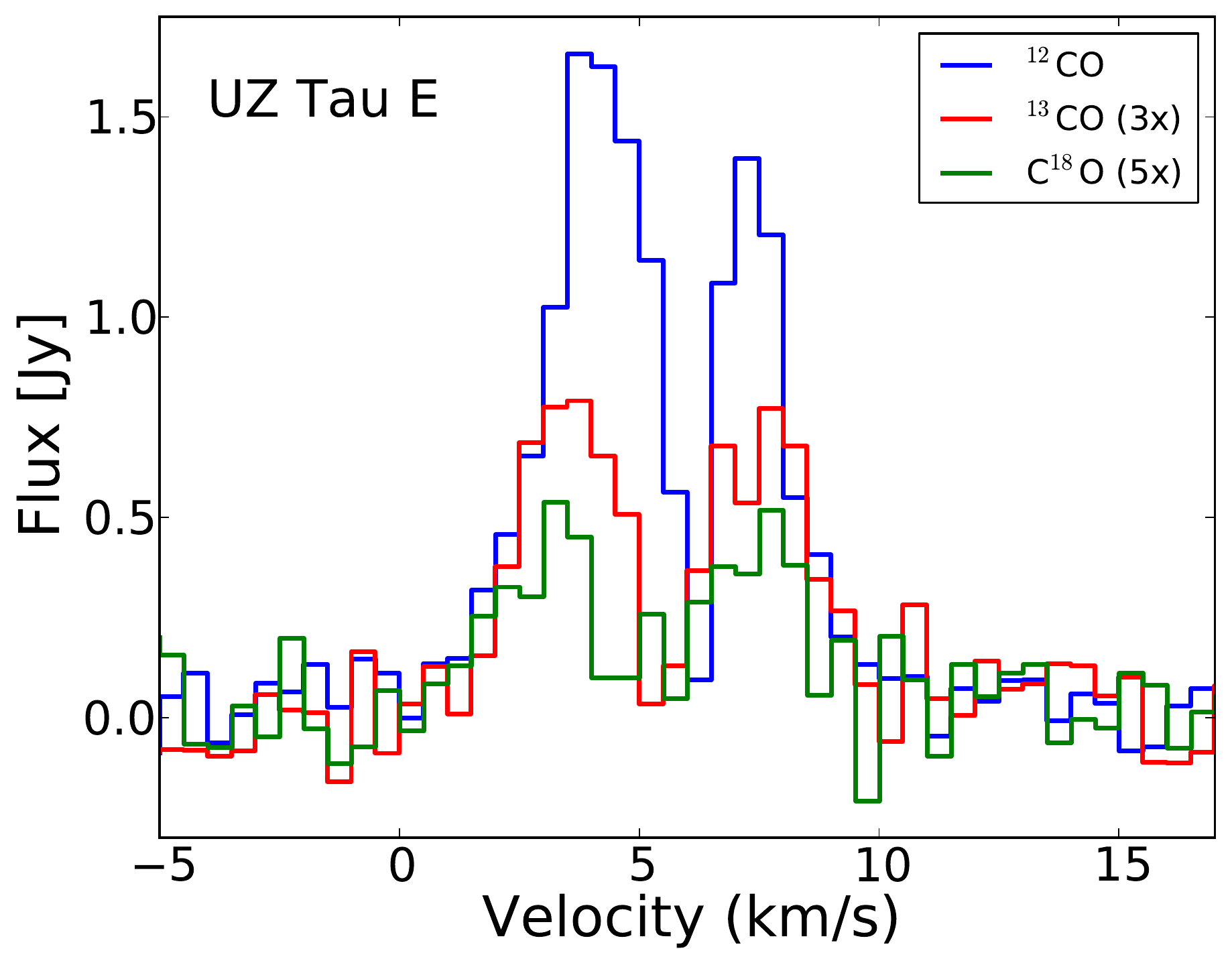}
  \includegraphics[angle=0,scale=0.3]{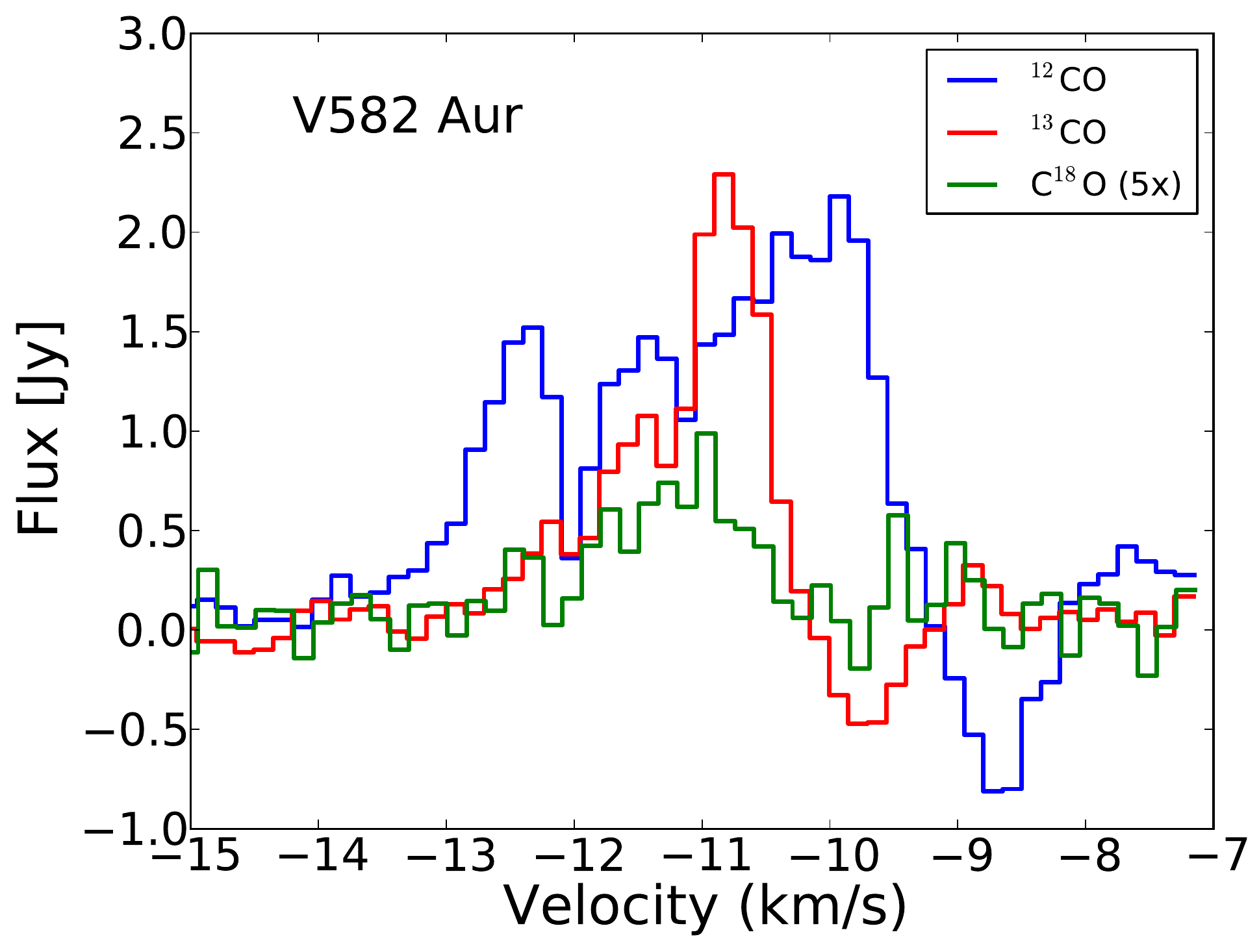}
  \caption{Integrated spectra for UZ~Tau~E and V582~Aur in $^{12}$CO,
    $^{13}$CO and C$^{18}$O.  The integrated line profiles were
    computed by integrating the emission in the regions corresponding
    to the dotted circles in Figure~\ref{mom0}. The spectra may have
    been scaled for display purposes, as specified in each figure.  }
\label{profiles}

\end{figure*}

\begin{figure*}
  \includegraphics[width=0.33\textwidth]{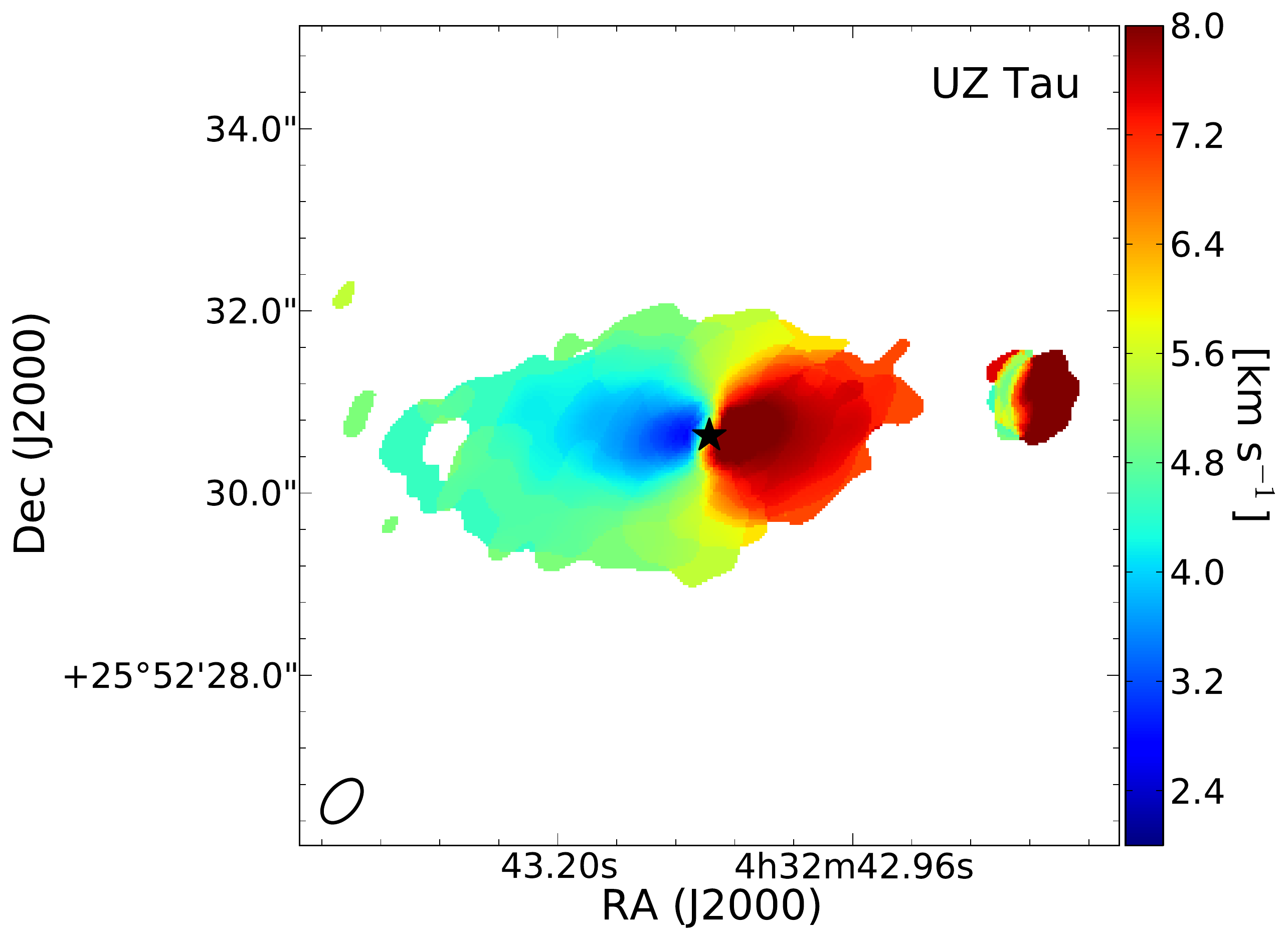}
  \includegraphics[width=0.33\textwidth]{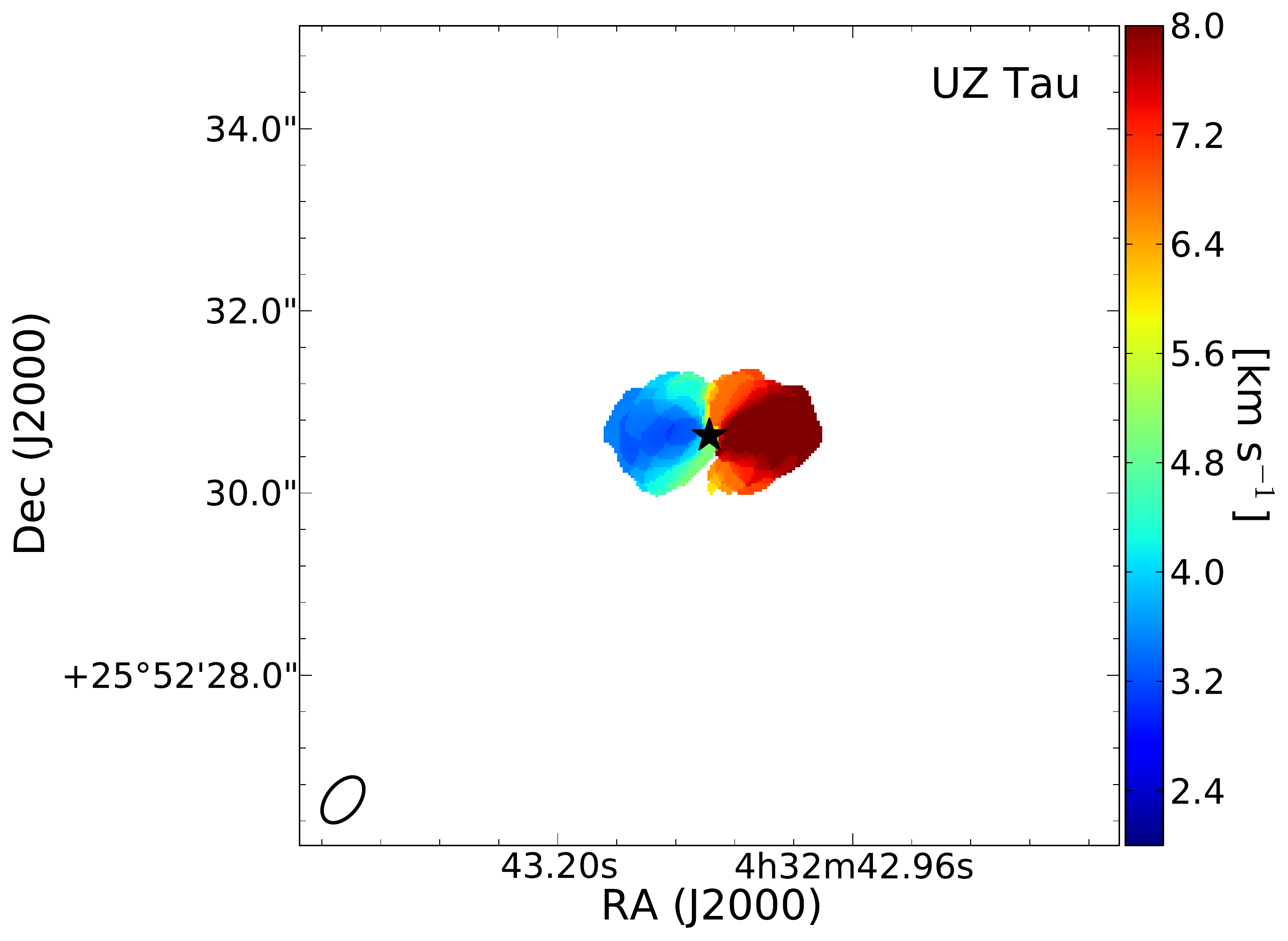}
  \includegraphics[width=0.33\textwidth]{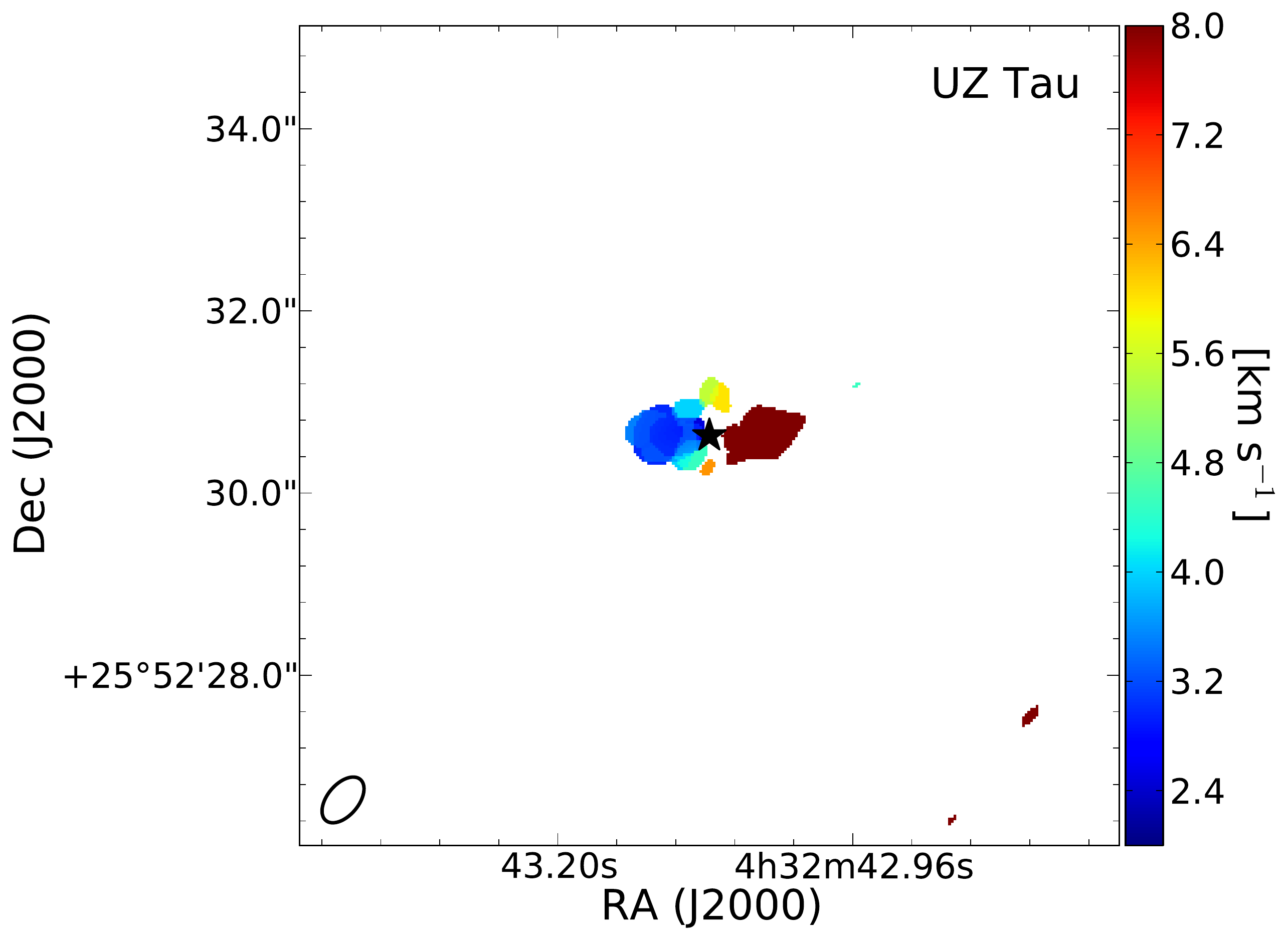}    
  \caption{Maps of projected velocity centroids (moment 1) in
    $^{12}$CO (left), $^{13}$CO (middle) and C$^{18}$O (right) for
    UZ~Tau~E.  }
  \label{mom1}
\end{figure*}

Integrated line emission (moment~0) maps for $^{12}$CO are shown in
Figure~\ref{mom0}. The total integrated line emission per molecule for
UZ~Tau~E and V582~Aur are presented in Table~\ref{cofluxes}.  The
moment maps and the final integrated line fluxes are all computed
integrating the channel with emission above 3-$\sigma$, and correspond
to the velocity ranges annotated in Figure~\ref{mom0}. The spectral
profiles for UZ~Tau~E and V582~Aur (integrated over the circular
apertures shown in Figure~\ref{mom0}) are shown in
Figure~\ref{profiles}. Figure~\ref{mom1} show intensity-weighted
velocity fields (moment~1) images in $^{12}$CO, $^{13}$CO and
C$^{18}$O for UZ~Tau~E. Velocity maps showing each CO isotopologue for
V582~Aur are presented in Appendix~\ref{co_maps}, together with moment~0
maps of $^{13}$CO and C$^{18}$O.

\begin{figure*}
  \includegraphics[height=0.34\textwidth]{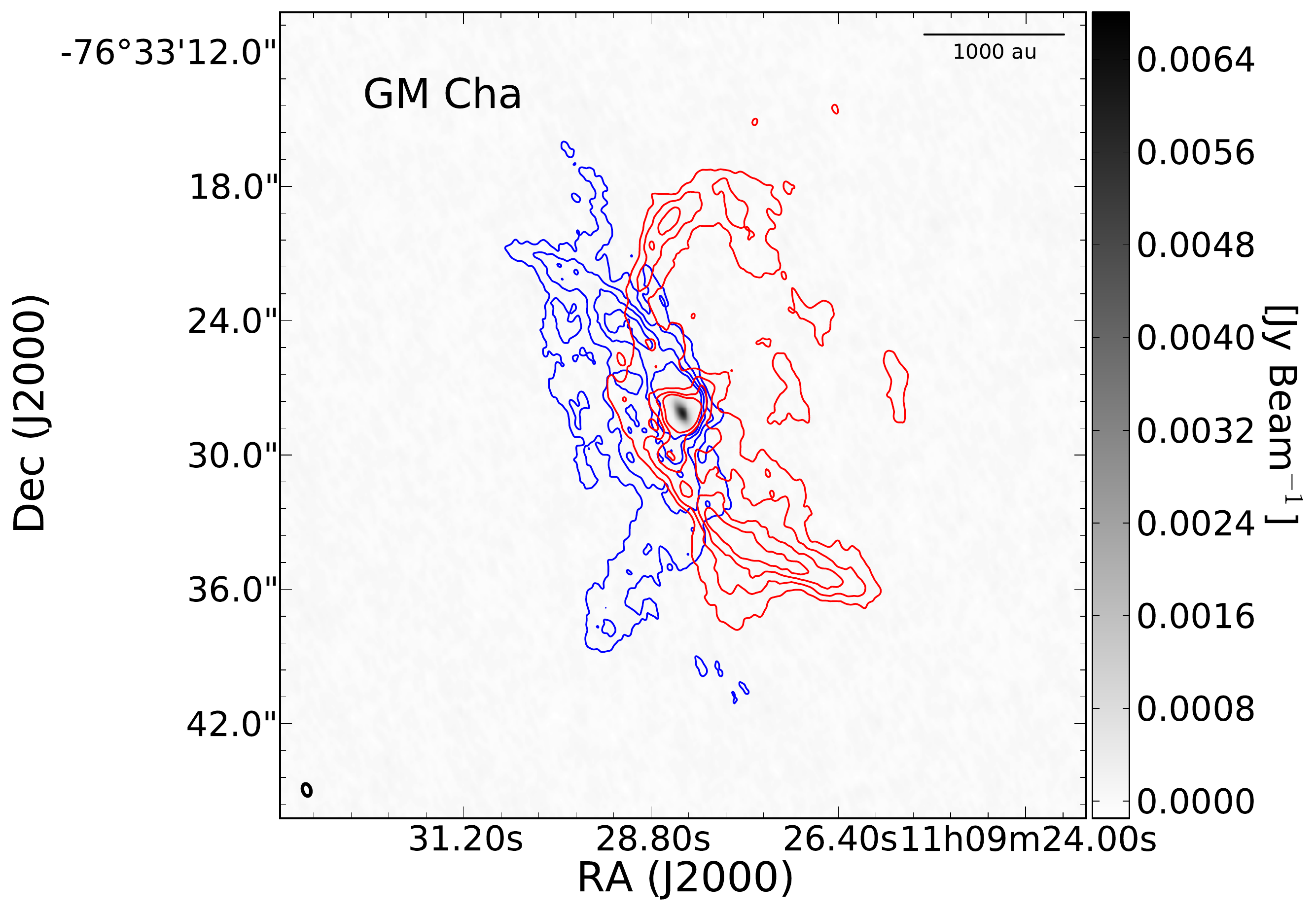}
  \includegraphics[height=0.34\textwidth]{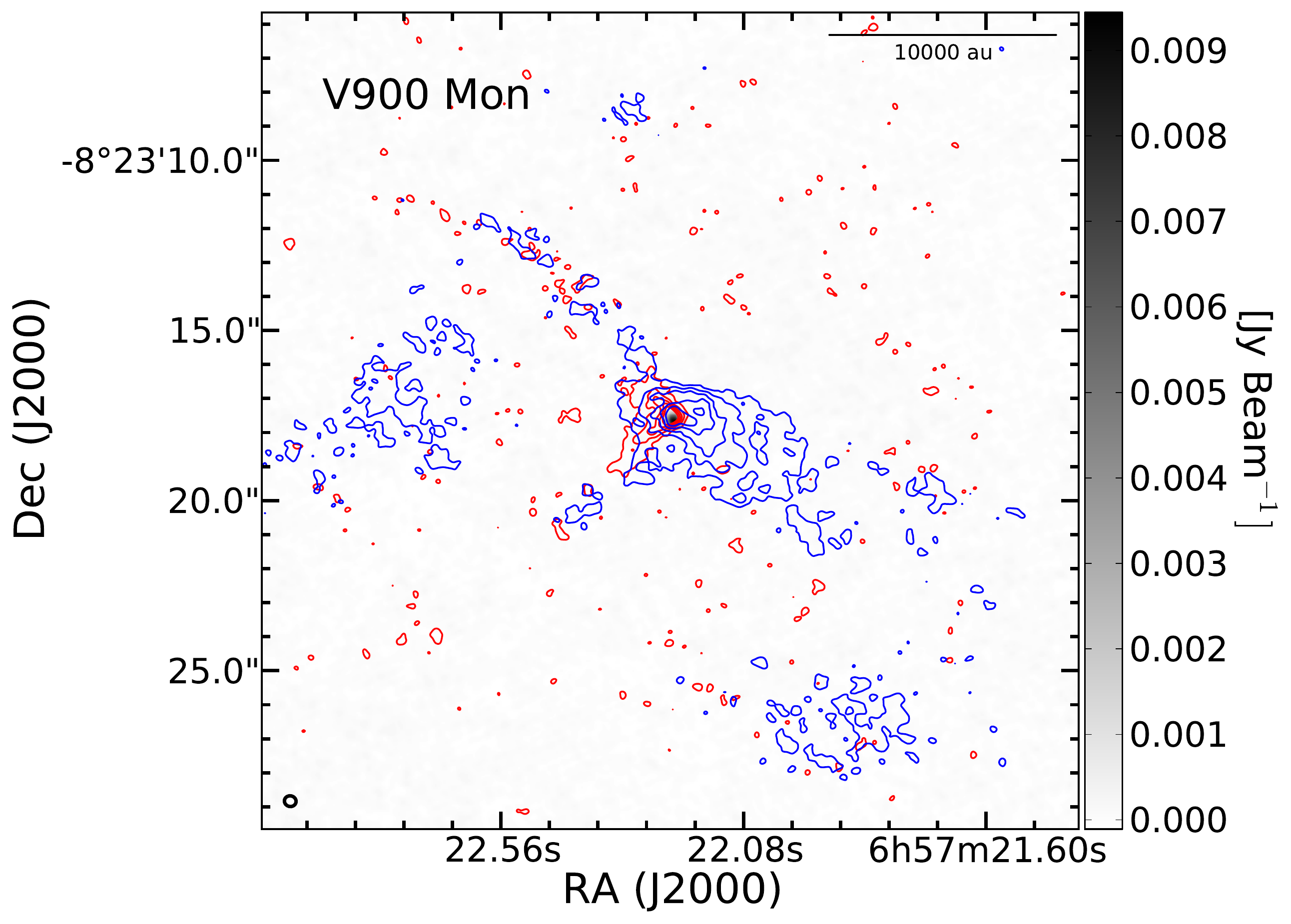}
          
\caption{{\it{Left: }}Blue and red contours show the integrated
  intensity of the $^{12}$CO blue- and red- shifted lobes of GM~Cha
  (integrated between -10 and 4.5~km~s$^{-1}$ and 6.5 and
  14.5~km~s$^{-1}$ respectively). Contour levels start at 3$\sigma$,
  increasing in steps of 2-$\sigma$.  The 1.3mm continuum image is
  shown in grayscale.  {\it{Right:}} Same for V900~Mon. Emission for
  the blue and red lobes was integrated between 13.5 and
  15.0~km~s$^{-1}$ and 15.5 and 18.0~km~s$^{-1}$ respectively. Contour
  levels start at 1.5$\sigma$, increasing in steps of 2-$\sigma$.  }
 \label{bluered}
\end{figure*}

All sources show very different morphologies in their
$^{12}$CO emission.  While the UZ~Tau~E disk shows a clear
Keplerian rotation pattern, V900~Mon and more noticeably GM~Cha both
show conical cavity walls similar to those detected around other FUors
(Figure~\ref{bluered}).  V582~Aur $^{12}$CO moment~0 map
  shows widespread emission and at least three peaks in its
  spectrum. This is due to its association to a star-forming
  filamentary cloud with velocities spanning
  $[-12.5,-7.5]$~km~s$^{-1}$ \citep{Dewangan2018,abraham2018}. The
  ALMA $^{12}$CO data shows compact emission within
  $\sim1.4$~arcsecond (3500~au) from the position of the FUor at
  velocities between $[-12.85,-11.5]$~km~s$^{-1}$ for $^{12}$CO, and
  $[-12.10,-11.5]$~km~s$^{-1}$ for $^{13}$CO, before cloud
  contamination becomes dominant (see full channel maps in
  Appendix~\ref{co_maps}). Central emission near V582~Aur in $^{13}$CO
  and C$^{18}$O was previously reported \citep{abraham2018} using
  NOEMA. Because of its location close to the source and velocities
  further away from the main cloud velocities, it is possible that
  this central emission is associated to the FUor.

\begin{deluxetable*}{lccc}
%\tablewidth{700pt}
\tablecaption{Measured CO J=2-1 Integrated Fluxes (angular integrated intensity)}
%\tabletypesize{\scriptsize}
\tablehead{
\colhead{Source} & 
\colhead{ $^{12}$CO} & 
\colhead{ $^{13}$CO} &
\colhead{C$^{18}$O}  \\
}
\startdata
UZ Tau & 7.02$\pm$0.02 & 1.16$\pm$0.02 & 0.42$\pm$0.01 \\
V582 Aur & 5.51$\pm$0.09 & 2.83$\pm$0.07& 0.81$\pm$0.035 \\
V900 Mon & 5.06$\pm$0.02 & 3.77$\pm$0.02 & 1.65$\pm$0.02 \\
GM Cha & 36.03$\pm$0.04 & 3.67$\pm$0.02 & 0.52$\pm$0.01  \\
\enddata
\tablecomments{Integrated line fluxes are in units of mJy~km~s$^{-1}$. Errors do not include the absolute flux uncertainty which is estimated to be between 7 and 10$\%$ in Band 6. }
\label{cofluxes}
\end{deluxetable*}

\section{Discussion}\label{discussion}

\subsection{Disk Masses}\label{masses}

A crude estimation of the disk dust masses can be obtained under the
standard assumption that the continuum emission from disks
  is optically thin at millimeter wavelengths
\citep[e.g.][]{Hildebrand1983}. This is certainly not true in the disk
inner regions (usually in the inner 5-10~au), but it is considered to
be a reasonable assumption for the rest of the disk \citep[e.g.,
][]{cieza2018}.

We estimate the disk dust masses from the optically thin assumption
that relates the observed flux, $F_\nu$, to the mass of solids
\citep[e.g.][]{beckwith1990}, via
  \begin{equation}
    M_{\rm dust} = \frac{F_\nu d^2}{\kappa_\nu B_\nu(T)},
  \end{equation}
\noindent where $B_\nu$ is the Planck function for a given temperature
$T$, $\kappa_\nu$ is the dust opacity and $d$ is the distance to the
emitting dust. We note that, at millimeter frequencies, the low
temperatures of some disks ($<$100~K) imply that the Rayleigh-Jeans
regime is not ideal for approximating the Planck function. For
example, at $T$=20~K and $\nu$=225~GHz, there is a 30\% difference
between calculating the full Planck function and the Rayleigh-Jeans
approximation. Therefore, caution must be taken when employing the
linear temperature dependence offered by Rayleigh-Jeans.

As discussed in \citet{cieza2018}, the assumption of optically thin
emission will underestimate the total dust mass, while the assumption
of a temperature of 20~K may overestimate the mass if the dust is
warmer, therefore the two effects may partially offset each other
(since the total mass approximately scales as the inverse of
temperature). To which extent this statement is valid may
  depend on the properties of each object. For instance, since FUor
  sources can be $\sim100$ times more luminous than normal stars in
  Class II disks, one would expect the average temperature of disks in
  FUor sources to be a factor of $\sim3$ higher than in normal Class
  II disks \citep[since the the disk temperature scales with the
    stellar luminosity T$\propto\,L^{1/4}$;][]{Chiang1997}. This will
  result in an overestimation of the disk mass by a similar amount.
  On the other hand, the more massive and/or compact become very
  optically thick in their inner regions and therefore the optically
  thin assumption will underestimate the total dust mass
  \citep[e.g.][]{liu2019}. In the case of V833~Ori, \citet{cieza2018}
  found that the total dust mass obtained using the optically thin
  approximation is similar to the mass derived using radiative
  transfer which takes into account the high optical depth in the disk
  inner regions. However, for HBC~494 and V2775~Ori, they found that
  the assumption of optically thin emission and 20~K dust temperature
  overestimate the mass by a factor of $\sim$2 when compared to the
  radiative transfer results. This can be attributed to the fact that
  the disks around HBC~494 and V2775~Ori are more compact than the
  disk around V833~Ori, and therefore the average disk temperatures
  are likely higher than 20~K. This highlights the need for combining
  resolved images with radiative transfer techniques in order to infer
  the true properties of these disks.

Here we use the dust opacity of \citet{beckwith1990},
  i.e. $\kappa_{\rm 1.3 mm}\approx 0.022$~cm$^2$~g$^{-1}$, to compute
  dust masses from the fluxes and distances listed in
  Table~\ref{tableobs}. We list calculations assuming two dust
  temperatures, 20~K and 60~K. The former is the typical temperature
  of a passively heated protoplanetary disks \citep{williams2011}
  while the latter aims to represent more active, hotter, disks \citep[see ][]{takami2019}.  The
derived disk dust masses are presented in Table~\ref{tableobs}. We
also list the total disk mass (gas+dust) by assuming the standard
gas-to-dust ratio of 100.

FUor sources in our sample have masses of solids which are at
  least one order of magnitude larger than those of the EXors,
consistent with the trend reported in \citet{cieza2018}. The assumed
temperature plays a significant role. Adopting a 3$\times$
  higher disk temperature yields a reservoir of solids
  $\sim$3.6 times smaller in mass.

The dust mass estimate for UZ~Tau~E is consistent with the value
derived by \cite{long2018} using higher resolution data at similar
frequency (67~M$_{\oplus}$ or
2.2$\times$10$^{-4}$M$_{\odot}$). Assuming a 100:1 gas-to-dust ratio
the total disk mass is 2.2$\times$10$^{-2}$M$_{\odot}$. The total disk
mass can also be obtained by comparing the $^{13}$CO(2--1) to
C$^{18}$O(2--1) integrated line ratios to the grid of models from
\citet{williams2014}. These models take into account basic CO
chemistry (photo-dissociation and CO freeze-out) and provide an
estimate of the total gas mass independent of the assumed gas-to-dust
ratio. The comparison to the models yield a total gas mass of
3.2$\times$10$^{-3}$M$_{\odot}$. Together with the dust mass estimated
above, this implies a gas-to-dust ratio of 14, which is similar to the
low ratios found around the EXor prototype EX~Lupi and other Class~II
disks around Lupus \citep{hales2018,ansdell2016,Miotello2017}. This
may be due to physical low amounts of gas, chemical conversion of CO
into other species, or other physical processes
\citep{Bosman2018,Krijt2018,Schwarz2018}.
  
We use radiative transfer codes in combination with MCMC methods to
infer the UZ~Tau~E disk parameters from the continuum and line data
independently. The dust and gas disks are modelled separately using
the radiative transfer code {\sc radmc-3d} \citep{Dullemond2012},
adopting the standard tapered-edge model to describe the surface
density profiles. See Appendix~\ref{uztaumodel} for details on the
modeling and MCMC procedure. These methods provide an alternative
estimation of the disk dust and gas masses.  We find that the masses
of the dust and gas disks are 92.9$^{+3.6}_{-13.3}$\mearth $\,$(2.8
$\times$ 10$^{-4}$\msun) and 7.8 $\times$
10$^{-4}$\msun\,respectively.

Both methods of deriving disk masses, from continuum and line
emission, have caveats. The continuum radiative transfer modeling only
accounts for a passive disk in hydrostatic equilibrium. While this is
arguably a decent approximation to a protoplanetary disk, it may be
far from the thermodynamical structure of actively accreting sources
whose energy budget should include extra energy terms such as viscous
heating, for example. This simplistic temperature structure also
affects the molecular gas line modelling. The line modeling
additionally suffers from uncertainties in CO isotopologue abundance
ratios and a limited accounting of photodissociation and freeze-out of
the CO molecules.  Bearing these caveats in mind, it is still possible
to use these masses to gain a rough idea of the gas-to-dust mass
ratio. This ratio would be approximately 2.8 for UZ~Tau~E, suggesting
its disk has a higher concentration of dust, or a lower amount of gas,
than the canonical assumption of a gas-to-dust mass of 100. The gas
mass may have been reduced due to disk-binary interactions \citep[see,
  for example,][]{Czekala2019}.

The total disk mass for V582~Aur, estimated assuming the dust emission
is optically thin (1055\mearth) and a gas-to-dust ratio of 100, is
0.3~M$_{\odot}$, which is a factor of 7.5 larger than the estimate of
\citet{abraham2018}.  This discrepancy is mostly due to the new GAIA
DR2 distance used in this work (2.5~kpc instead of 1.3~kpc), and
cooler dust temperature (20~K instead of 30~K).

\begin{figure*}
  \begin{center}
    \includegraphics[angle=0,scale=0.55]{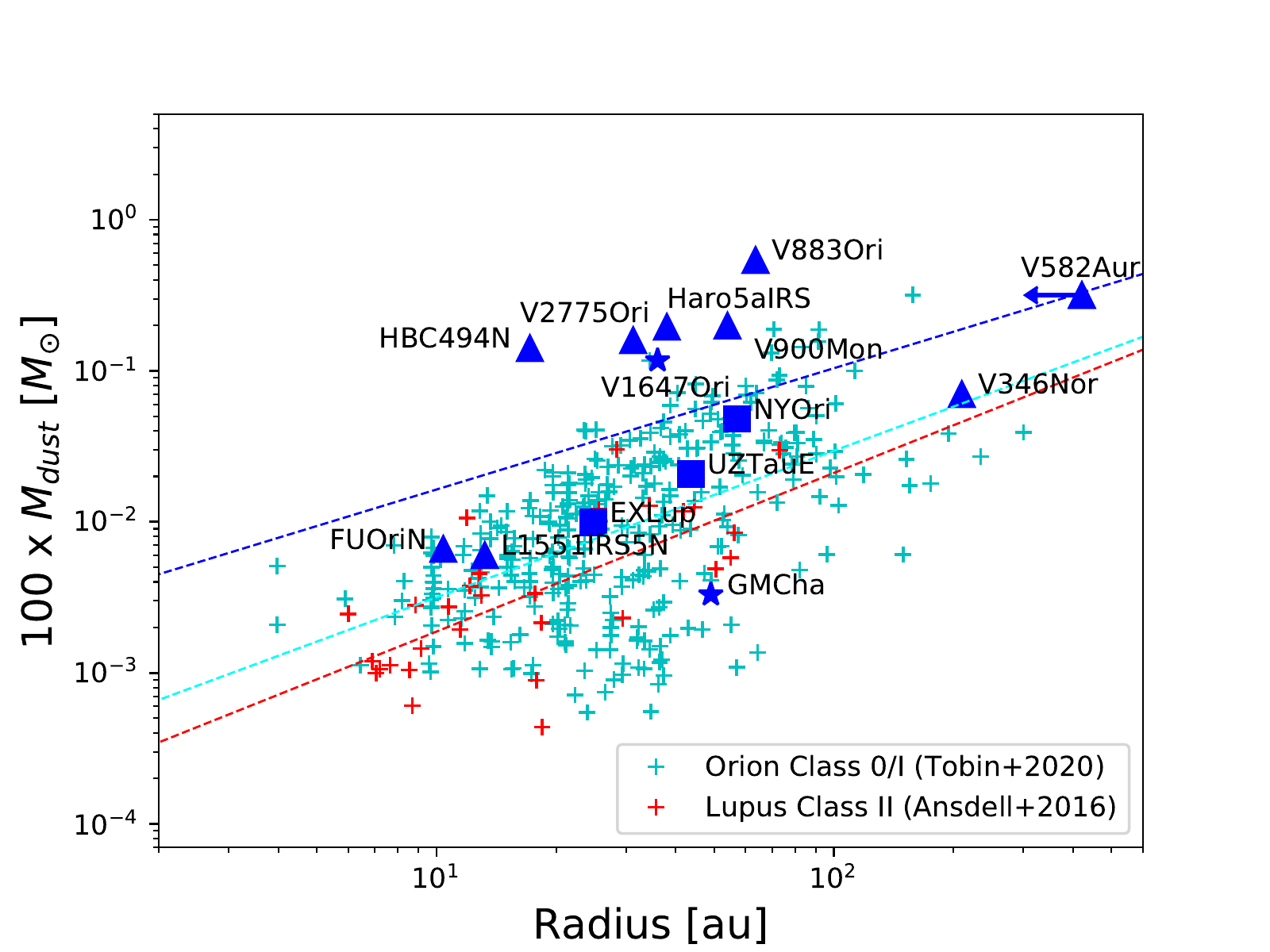}%{0.4\textwidth}
    \includegraphics[angle=0,scale=0.55]{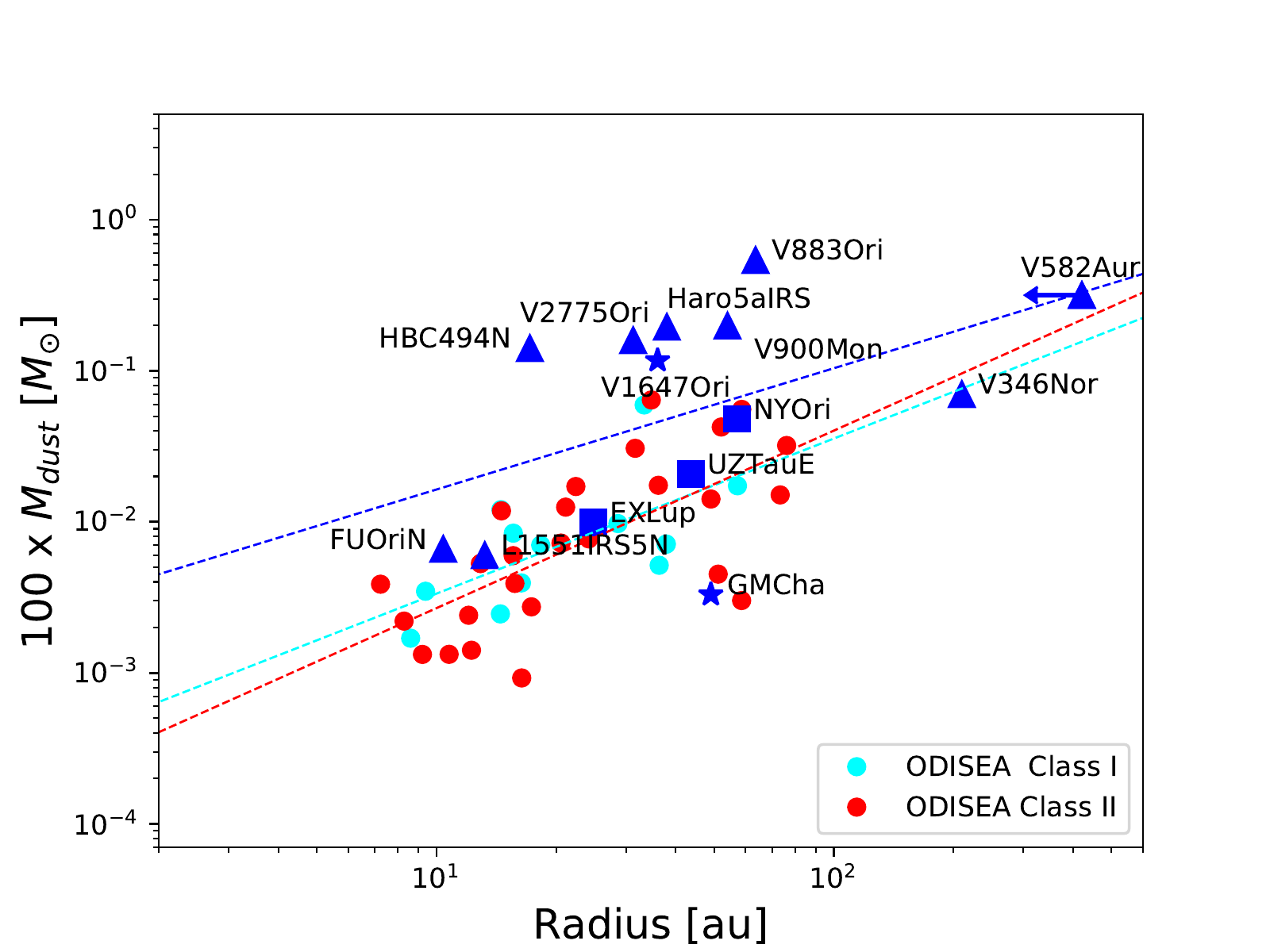}%{0.4\textwidth}
    
    \caption{ Relationship between disk mass and disk radius (FWHM/2)
      for FUor/EXor sources presented here (assuming a 20~K dust
      temperature) and in the literature. Squares are EXors, triangles
      are FUors, and objects with double FUor and EXor classification
      are marked with a star symbol. The disk size and mass for HBC494
      correspond to the ones measured towards the brightest component
      of the binary, HBC494 North \citep{zurloprep}. The horizontal
      arrow on V582~Aur denotes the disk radius is an upper
      limit. {\bf{Left panel}}: Light blue crosses show the disk
      masses and radii (FWHM/2) for Class 0 and I protostellar disk
      candidates in Orion derived from 0.87~mm ALMA observations
      \citep{tobin2020}. Red crosses show the disk masses and disk
      radii for Class~II sources in Lupus from
      \citet{ansdell2016}. The dashed lines correspond to power law
      fits to the Class 0/I, Class II and FU/EXor data respectively.
      {\bf{Right panel}}: Same as left panel, but now comparing the
      FU/EXor data to Class I and Class II sources in Ophiuchus from
      the ODISEA sample \citep{cieza2019}.}
    \label{lucas}
\end{center}
\end{figure*}

\subsection{Disk Sizes}\label{Sizes}

Figure~\ref{lucas} shows an updated version of Figure~6 in
\citet{cieza2018}, in which disk sizes and total disk masses from the
FU/EXor sample are compared to those of other protostellar and
protoplanetary sources observed by ALMA \citep[][ and also this
  work]{kospal2017b,cieza2018,hales2018,kospal2018,takami2019,cruz2019,perez2020}.
The properties of Class 0 and I protostellar disk candidates in Orion
\citep{tobin2020} and the Class II disks in Lupus \citep{ansdell2016}
are shown for comparison.  We also compare the properties of eruptive
sources to those of protostellar and protoplanetary disks in Ophiuchus
from the Ophiuchus DIsk Survey Employing ALMA \citep[ODISEA,
][]{cieza2019}, which currently contains 12 Class I and 26 Class II
sources with measured 1.3mm fluxes and resolved disk sizes.

The disk masses for the ODISEA sample are estimated assuming an
optically thin dust continuum, a dust temperature of 20~K, and a
gas-to-dust ratio of 100. Disk sizes are estimated from the
deconvolved Gaussian FWHM/2 radius obtained from 2D Gaussian fits to
the continuum images.  \citet{cieza2018} used the characteristic
radius R$_{\rm c}$ (see Appendix \ref{uztaumodeldust}) as proxy for
disk radius, while other authors have used the Gaussian 2$\sigma$
radius or the curve-of-growth method to estimate the radius that
contains a certain fraction of the total disk mass \citep{bate2018} or
a fraction of the total flux
\citep{tripathi2017,Ansdell2018,Trapman2019}. There are advantages and
caveats on the ability of each method to measure the true sizes of
circumstellar disks, which is beyond the purpose of this work
\citep[see discussions in e.g.][]{tripathi2017,bate2018,tobin2020}. We
choose to use the deconvolved Gaussian FWHM/2 radius since it provides
an homogeneous measure across the selected ALMA datasets.\\ We find
that the new eruptive sources presented here follow the same trend as
the sample in \citet{cieza2018}, in which FUor disks are more massive
than Class II sources of similar size. In contrast, the properties of
the EXor disks are more consistent with those of Class II objects. As
noted by \citet{tobin2020}, there does not seem to be a clear
distinction in disk sizes between Class II in Lupus and the younger
Class 0/I in Orion. Interestingly, we also find no difference when
comparing the Class II disk sizes to the Class I sources in Ophiuchus.
This suggests that for a given disk size, the disks around FUor disks
are brighter than disks of all protostellar classes. EXor disks, on
the other hand, do not seem to be different from non-eruptive disks.

Two FUor disks deviate from this trend, FU~Ori~North and
L1551~IRS~5~North, both of them are binary systems. The class
prototype FU~Ori~North hosts the smallest disk of all eruptive targets
\citep{perez2020,liu2019}, which is part of a 210~au separation binary
system. Both disks in the FU~Ori system have similar 1.3mm
sizes. \citet{cruz2019} detected compact disks of similar sizes in
another binary FUor-like system L1551~IRS~5. Recently
\citet{zurloprep} resolved for the first time two binary components
and their respective disks in the HBC~494 system which are also
compact ($<$20~au). The small disk sizes found around binary FUor
systems are consistent with observations of non-eruptive multiple
systems in Orion.  \citet{tobin2020} found that multiple systems show
the smallest (and faintest) distribution of disk sizes, which indicate
that multiplicity has a significant impact on disk evolution. The disk
sizes measured around binary FUor systems are consistent
with simulations of stellar fly-bys in binary systems by
\citet{Cuello2019}.  These simulations show that stellar
encounters can deplete the disk outer regions and increase inner disk
density, which can in turn enhance stellar accretion and explain the
accretion outbursts of binary FUor systems.\\

The power law fits in Figure~\ref{lucas} can be inverted to estimate
the relation of disk radius with disk mass. \citet{tobin2020} reports
a power-law relation between disk radius and mass of R$\propto$
M$^{0.30 \pm 0.03}$, which is similar to the power law scaling
predicted by models of disk formation regulated by magnetic fields
(which depend on magnetic field strength, ambipolar diffusion
timescale and combined disk+stellar mass). The power law dependence
for Class II in Lupus is steeper (R$\propto$M$^{0.46 \pm 0.09}$),
closer to the square-root dependence predicted for optically thick
disks. The fit to the FUor sample indicates that disk radii increase
with mass as R$\propto$ M$^{0.27 \pm 0.15}$ . This is closer to the
value of 0.3 found for Class 0/I protostars in Orion although with a
larger uncertainty dominated by low number statistics. Increasing the
FUor sample observed at sufficient high angular resolution will help
improving the accuracy of the fit and allow a more consistent
comparison with larger samples of Class 0/I and Class II sources. The
size-luminosity relation has been studied intensively around over 200
Class II disks, for which a correlation between millimeter luminosity
and dust disk size of L$\propto\,R^2$ has been well characterized
\citep[in agreement with the R$\propto$M$^{0.46}$ found for the
  Class~II in Lupus, since millimeter luminosity is a proxy for
  mass;][]{tripathi2017,andrews2018}. \citet{Rosotti2019} showed that
the size-luminosity relation can be explained by grain growth and
radial drift, although it could also be related to disk initial
conditions or to optical depth effects \citep[][ and references
  therein]{andrews2010,andrews2020}. Measuring the size-luminosity
relation in episodically accreting disks, which are thought to be a
common early phase of evolution of disks around low mass stars, may
inform on the initial unruly dynamics of protoplanetary disks. If
these active disks indeed reveal an earlier stage in disk evolution,
their size-luminosity relation allows for a more realistic description
of the initial condition for later disk models, as an alternative for
the classical 'steady state' start.

\subsection{Gas Kinematics and Morphology} \label{gas}

Complex kinematics can be the product of several types of dynamical
interactions, such as flybys and binary interactions, outflows and
winds, embedded planets and vortices, and the capture of cloud
material. Several active sources are known binaries, such as UZ~Tau~E
for example, and recent hydrodynamic models of tidal interactions have
shown that tidal encounters have distinct kinematics signatures
\citep{Cuello2019} and that these encounters may be related to
outbursting events. Interactions with an embedded sub-stellar
perturber can excite kinematic perturbations \citep{perez2015} which
could persist of large scales in the disk \citep{perez2018}. On the
other hand, \citet{Dullemond2019} recently speculates that outbursting
systems may be the result of cloudlet capture events, which do bear
observational kinematics in the form of complex tail-like features in
maps of line emission.

The molecular line data for the four sources observed have distinctive
morphological features. GM~Cha and V900~Mon $^{12}$CO(2--1) integrated
maps show the presence of wide-angle conical cavities. The presence of
an $^{12}$CO(3-2) outflow around GM~Cha was first reported by
\citet{Mottram2017} using APEX.  The blue-shifted and red-shifted
lobes are more clearly distinguished in GM~Cha, while in V900~Mon the
distinction is less clear. These morphologies are similar to the one
observed in V883~Ori, HBC~494, V2775~Ori and V1647~Ori
\citep{ruiz2017a,ruiz2017b,zurlo2017,principe2018}, in which opening
cavities in $^{12}$CO(2--1) are reported around all FUor candidates
but not around the EXors. The observed emission is interpreted as the
walls of a cavity that is carved out by a slow moving outflow,
probably produced by material swept-up by a faster jet bow-shock
\citep{Frank2014}. As the sources evolve from Class I to Class II,
outflows carve out cavities that widen as the source ages
\citep{arce2006}, in a process that results in the dispersion of the
remaining pre-stellar core.  Combining ages derived using the L$_{\rm
  Bol}$-age relation from \citet{Ladd1998}, and the measured opening
angles, \citet{arce2006} found that sources with opening angles larger
than $\sim$150 degrees have ages of $\sim$10$^6$~yr, closer to the
typical ages of Class~II stars.

The opening angles for the blue- and red-shifted lobes of GM~Cha are
both very similar $\sim$120~degrees, measured at a distance of
1300~au).  For V900~Mon the opening angle measured on the
blue-shifted lobe is $\sim$90~degrees.  The opening angles of GM~Cha
are comparable to wide opening angles of 150~degrees measured around
HBC~494 and V883~Ori \citep{ruiz2017a,ruiz2017b}. As noted by
\citet{takami2019}, the opening angle of $\sim$V900~Mon is narrower
than the one around these sources, but wider than the one measured
around the eruptive Class~0 protostar V346~Nor \citep{kospal2017b}.
  
The presence of outflows in FUor sources \citep{ruiz2017a, ruiz2017b,
  kospal2017b, zurlo2017} and not around EXors suggests a distinction
in the evolutionary stages between the two classes, with FUors closer
to Class~I sources. Nevertheless, EX~Lupi and the ambiguously
FUor/EXor classified V1647 Ori also have outflows
\citep{principe2018,hales2018}. The outflows around these two EXors
are fainter than the ones around FUors, therefore the distinction is
more subtle and supports the idea of a continuous evolution from FUor
to EXor, with EXors more evolved than FUors.

V582~Aur does not show a clear opening cavity but this may be due to
the greater distance to the source. As a point of reference, V900~Mon
is two times closer than V582~Aur but its outflow cavity walls were
barely detected above the noise in our data. Another possibility is
that the optical variability in V582~Aur is due to variable extinction
and not accretion outbursts \citep{abraham2018}.  This would reinforce
the picture that prominent cavity walls are carved during strong
accretion outbursts. Deeper, higher resolution images of the more
compact emission observed at the V582~Aur position could provide
better understanding of its kinematics and relation to the FUor.

UZ~Tau~E, the only bona-fide EXor from our sample, is noticeably
different from the FUors. It shows the clear Keplerian pattern of a
Class II disk. This supports the evolutionary distinction between
EXors and FUors.  Cloud or envelope emission are most likely
responsible to why keplerian rotation is difficult to observe around
most FUors. $^{12}$CO and $^{13}$CO are typically too contaminated by
envelope emission, but a Keplerian disk is clearly seen in C$^{18}$O
in V883~Ori \citep{cieza2016}.  It is possible that Keplerian rotation
is not detected in some FUors simply because the observations are not
deep enough, as C$^{18}$O is generally too faint for easy detection in
other lower-mass stars \citep{Ansdell2018}. \citet{Ansdell2018}
detected C$^{18}$O in the most massive disks in their sample, however
most FUors are located 2-3 times further away than Lupus, with more
extreme cases such as V900~Mon and V582~Aur that are 10 and 15 more
distant than Lupus, respectively).  Another possibility is that FUor
disks are too perturbed and their kinematics distorted \citep[due to
  the potential disk precession;][]{principe2018}. High resolution
observations of FU~Ori reveal a rotation pattern distorted by complex
kinematics \citep{perez2020} possibly tracing binary interactions near
the base of the outflow.

\subsection{Triggering mechanisms}

Observations of FUOr objects suggest that the FUor phenomenon is an
heterogenous process. Some systems harbour large, massive disks such
as the ones detected around V883~Ori, V2775~Ori and V582~Aur.  High
resolution millimeter observations have also identified binary systems
with compact, hot disks in each of the components (like FU~Ori itself,
HBC~494, and L1551~IRS~5).

Gravitational instability requires that the mass of the disk is at
least 10\% the mass of the central star in order to operate \citep[see
  discussion in ][and references therein]{cieza2018}, and could
possibly explain the outburst in the most massive disks. However high
resolution observations do not show the signatures predicted by
theoretical simulations of gravitationally unstable disks, as was
demonstrated in the case of V883~Ori. High resolution ALMA images did
not reveal the spiral or clumpy features predicted by GI simulations
\citep{cieza2016}. V582~Aur hosts the most massive disk our sample,
yet being located 2.5~kpc away it was unresolved by our observations
and therefore we are unable to test the GI scenario. Nevertheless,
none of the massive FUor disks that have been resolved so far have
show these signatures, which at least rules the possibility of GI
operating in spatial scales compared to the spatial resolution of the
observations. As pointed out by \citet{cieza2018}, the lack of large
scale instabilities of fragmentation in these observations suggest
that the outburst of these systems could support models that combine
MRI and GI without fragmentation \citep[e.g. ][]{zhu2009b}.

The outbursts in binary systems could be explained by perturbation of
the accretion disk by a stellar companion. \citet{Bonnell1992} showed
that binary perturbation can render the disk unstable and could
possibly explain the triggering of FUor outbursts in multiple
systems. As mentioned earlier, stellar fly-by scenario could also
explain the compact sizes of the disks in binary systems as
simulations by \citet{Cuello2019} show that an inclined prograde
encounter can remove material from the outer disk and increase the
inner disk density (which in turn can enhance the
accretion).

Observations of EXors reveal disks that are similar to those observed
around typical Class II sources, both in size and mass. The disks are
not massive enough to satisfy the requirement for triggering
gravitational instability. Most EXors resolved by ALMA so far, with
the exception of UZ~Tau~E, are single star systems and therefore there
is no clear connection between the EXor phenomenon and stellar
multiplicity. As pointed out already by \citet{cieza2018}, the
triggering mechanisms for EXor outburst is more likely associated to
instabilities in the inner disks and/or interactions between the disk
and planetary companions \citep{lodato2004}.

\section{Conclusion}\label{conclusion}

We conducted a campaign to observe four young eruptive stellar systems
with ALMA at 0\farcs4 resolution.  This sample represents 10\% of
known FUor/EXor objects in the non-exhaustive list of eruptive young
stars from \citet{audard2014}, increasing the number of eruptive
sources observed at sub-arcsecond resolution by 15\%. We detected
1.3mm continuum emission in all four sources.

We found that the FUors have dust disks that are more massive than
those around found around Class 0/I sources and Class~II objects of
similar size, making them more likely to become gravitationally
unstable and trigger the outburst.  The EXor in our sample has a dust
and gas disk that is well modeled with a passively irradiated disk in
hydrostatic equilibrium, similar to those found in Class II sources.

We find that two of the three FUor objects show prominent outflows in
molecular gas emission.  While the FUor V900~Mon shows a distinct
conical cavity similar to those of Class~I objects, the source with
unclear FU/EXor classification (GM~Cha) has a wide-angle outflow
similar to those found around a subset of FUor objects such as
V883~Ori and HBC~494 \citep[and FU Orionis itself;
][]{halesprep}. Although the sample size remains small, the presence
of outflow activity in FUors and not in EXors suggests that the two
types of objects represent different evolutionary stages, with EXors
more evolved than FUors.

These results highlight the importance of spectral line observations
sensitive to various spatial scales for inferring the nature of
eruptive sources, which seem to span from Class~0 to the early Class
II stages of protostellar evolution. Observations targeting larger
scale structure ($>$1000~au) are required, to determine the properties
of outflows around FUors (and confirm their absence in EXors), while
deeper continuum and C$^{18}$O observations at few au resolution will
measure the properties of their inner disks and help understand what
drives this important early phase of star formation.

\software{Common Astronomy Software Applications
  \citep{2007ASPC..376..127M}, {\sc radmc-3d}, \citep{Dullemond2012},
  GALARIO, \citep{tazzari2017,tazzari2018}, EMCEE \citep{foreman2013},
  Astropy \citep{2013A&A...558A..33A}}

\section*{Acknowledgments}

We thank the anonymous referee for a very constructive report. S.P
acknowledges support from ANID-FONDECYT grant 1191934 and from the
Joint Committee of ESO and the Government of Chile.  A.Z. acknowledges
support from the FONDECYT Iniciaci\'on en investigaci\'on project
number 11190837.  This paper makes use of the following ALMA data:
ADS/JAO.ALMA\#2017.1.01031.S. ALMA is a partnership of ESO
(representing its member states), NSF (USA) and NINS (Japan), together
with NRC (Canada) and NSC and ASIAA (Taiwan), in cooperation with the
Republic of Chile. The Joint ALMA Observatory is operated by ESO,
AUI/NRAO and NAOJ. The National Radio Astronomy Observatory is a
facility of the National Science Foundation operated under cooperative
agreement by Associated Universities, Inc. This work has made use of
data from the European Space Agency (ESA) mission {\it Gaia}
(\url{https://www.cosmos.esa.int/gaia}), processed by the {\it Gaia}
Data Processing and Analysis Consortium (DPAC,
\url{https://www.cosmos.esa.int/web/gaia/dpac/consortium}). Funding
for the DPAC has been provided by national institutions, in particular
the institutions participating in the {\it Gaia} Multilateral
Agreement.

{}

\appendix

\section{Radiative Transfer Modeling of UZ~Tau~E}\label{uztaumodel}

\subsection{Dust continuum model}\label{uztaumodeldust}

The millimeter continuum disk around UZ~Tau~E has been studied
intensively in previous works: \citet{tripathi2018} using CARMA,
\citet{long2018} and \citet{manara2019} using higher resolution ALMA
data. To compare the properties (mainly size and perhaps the mass) of
the dust disk around UZ~Tau~E to the ones of other eruptive sources
studied by our group, we modelled the continuum emission with a simple
passively heated disk. The model assumes a power-law for the surface
density profile with an exponential taper beyond a characteristic
radius.  The exact formulation of the model and the radiative transfer
calculation are described in \citet{cieza2018}, \citet{hales2018} and
\citet{perez2020} --we refer the readers to these works for a detailed
description of the radiative transfer modelling.

The disk model is described by 5 free parameters: the dust mass
M$_{\rm disk}$, the slope of the surface density power law $\gamma$,
the characteristic radius R$_{\rm c}$, the scale height at 100 au
H$_{\rm 100}$ and the flaring index $\Psi$. The flux emerging from the
parametric disk model is computed using the radiative transfer code
{\sc radmc-3d} \citep[version 0.41,][]{Dullemond2012}.

The stellar parameters that we adopt for UZ~Tau~E are: an effective
stellar temperature of $3574$~K and a luminosity of 0.35~L$_\odot$
\citep{long2018}. The assumed distribution of dust grains and their
optical properties are the same as the ones adopted in our previous
works (the dust absorption opacity at 1.3 mm is $\kappa_{\rm abs} =
0.022$~cm$^2\,$g$^{-1}$). The inclination angle $i$ and PA of the
model are fixed to the values derived by \citet{long2018} from higher
resolution images.

The {\sc emcee} MCMC algorithm \citep{foreman2013} was used to sample
the posterior distributions of each parameter. We run RT models with
240 walkers for 1000 iterations. The resulting posteriors are shown in
Fig. \ref{fig:fit}. We note that since the model is a passive disk (it
neglects,for example, viscous heating), the derived temperatures serve
only as a crude approximation for the dust temperature. Nevertheless,
the best-fit (maximum likelihood set of parameters) model allows to
obtain a crude estimate of the size and bulk mass of the disk. The 1.3
mm observations of UZ~Tau~E can thus be described by a disk profile
with characteristic radii of 61 au and total dust mass of 92.9~\mearth
(2.8$\times$10$^{-4}$\msun). The slope of the surface density
distribution is 0.8, similar to those of T~Tauri stars
\citep{andrews2010}.  See corner Figure~\ref{fig:fit} for parameter
uncertainties. The parameters' posterior distributions show that there
could be some degeneracies in our modelling. We think this is could be
due to underlying structures not accounted for in our simple
model. Higher resolution data, and a more complete model that includes
viscous heating, would be required to describe the UZ~Tau~E disk in
detail.

\begin{figure}
  \begin{center}
    \epsscale{0.5}
    \includegraphics[angle=0,scale=0.5]{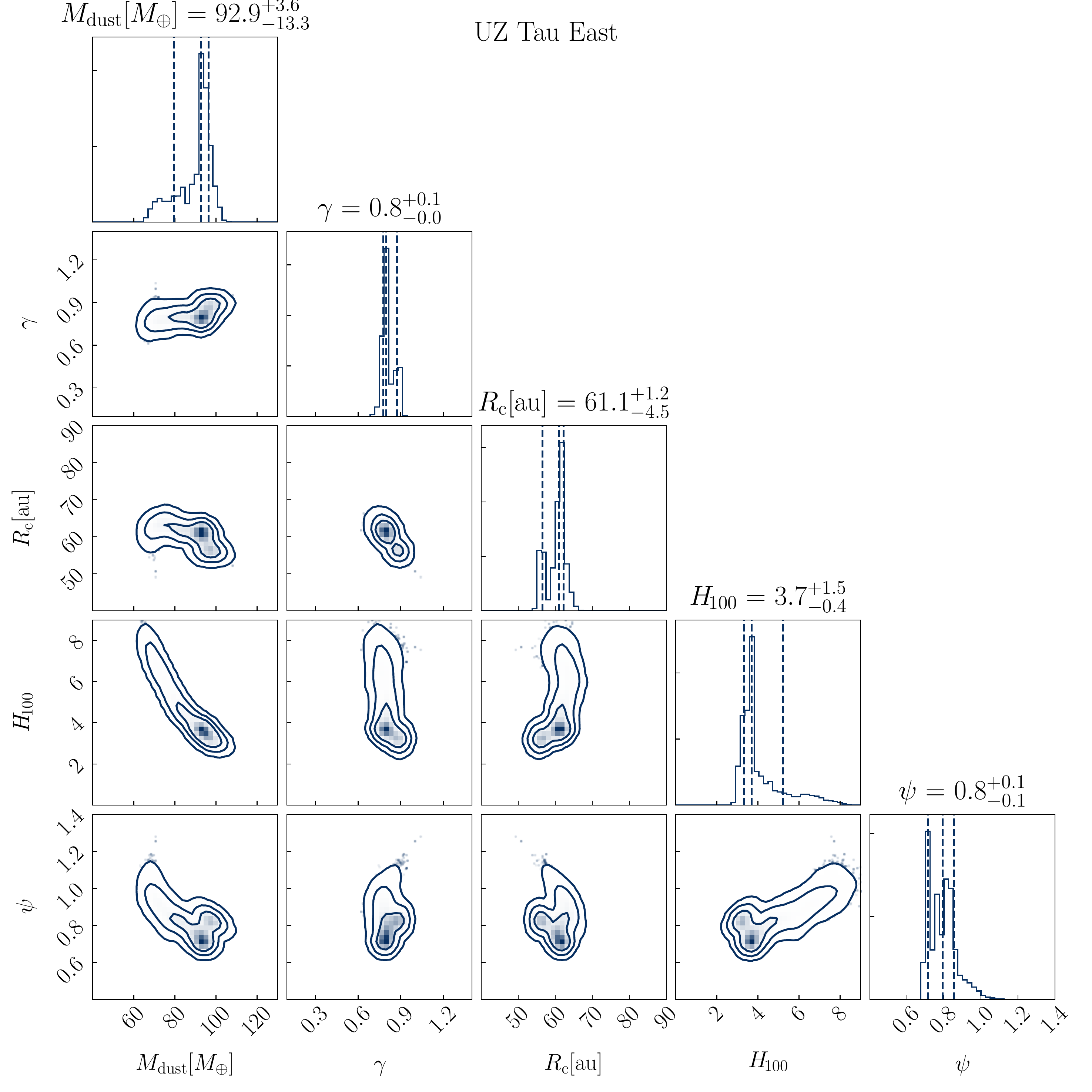}
    \caption{Triangle plots of the posterior probability distribution function for the different model parameters.}  
    \label{fig:fit}
  \end{center}
\end{figure}

\section{$^{13}$CO and C$^{18}$O gas model}\label{uztaumodelgas}

In this section we focus in modelling of the $^{13}$CO and C$^{18}$O
gas emission detected by our ALMA observations. We use the {\sc pdspy}
code from \citet{Sheehan2019} to generate synthetic molecular line
emission channel maps from a model of a Keplerian-rotating, passively
irradiated disk in hydrostatic equilibrium. The models are then used
to fit the $^{13}$CO and C$^{18}$O data simultaneously using a Markov
Chain Monte Carlo (MCMC) routine. The code is described in detail in
\citet{Sheehan2019}, and it uses {\sc RADMC-3D} to ray-trace the
density and temperature structure, and {\sc GALARIO} for fast sampling
of the visibilities from the synthetic channel maps generated by
RADMC-3D \citep{tazzari2018}. The model visibilities are then provided
as input to an MCMC routine that makes use of the {\sc emcee} code
\cite{foreman2013} to compare the synthetic observations with our data.

The model adopted here includes 13 parameters: total disk mass M$_{\rm
  disk}$, combined stellar mass M$_{*}$, disk characteristic radius
R$_{\rm disk}$, disk inner radius R$_{\rm in}$, T$_0$, a$_{\rm turb}$,
position angle PA, $q$, system velocity v$_{\rm sys}$ (LSRK), and
offset from the phase center x$_0$ and y$_0$. The dynamical mass of
the central object M$_{*}$ is fitted as part of our model, assuming
the distance of 131.2$\pm$1.7~pc \citep{gaia2018}. The disk is assumed
to be vertically isothermal, with a power-law radial dependence of the
temperature given by
\begin{equation}\label{eqn7}
T(r)=T_0\Big(\frac{r}{1au}\Big)^{-q},
\end{equation}
where T$_0$ corresponds to the temperature at 1~au, and $q$ is the
radial exponent for the temperature dependence. The surface density
profile of the disk is given by Equation A1, with the vertical
structure determined by using the vertically isothermal temperature
(Equation B4) to solve for hydrostatic equilibrium.  Following
\citet{Czekala2019} we fixed $q$ to 0.5, the surface density power law
exponent $\gamma$ to 1.0 and the inclination angle of the disk to
56.16$^\circ$.  M$_{\rm disk}$ is the total gas mass. The abundance of
$^{12}$CO relative to H$_2$ was set to 10$^{-4}$ and assumed constant
throughout the disk. We adopt the canonical values for the
  $^{13}$CO and C$^{18}$O isotopologue ratios with respect to
  $^{12}$CO \citep[77 and 550, respectively;][]{Wilson1994}.

Each run was started with 100 walkers and were run for $\sim$2500
steps. Detailed description of the MCMC fitting procedure can be found
in Appendix~A1 of \citet{Sheehan2019}. In Table~\ref{fit} we show the
best fit parameters obtained computing the median value of the MCMC
samples after removing the burn-in steps. The triangle plots of the
posterior probability distribution function from the MCMC fitting
process are shown in Figure~\ref{fig:fit}. The uncertainty for each
variable parameter are estimated by considering the range around the
median value that contains 68\% of the walker positions (after
removing the burn-in steps).

We find a combined stellar mass for M1+M2 of
1.25$\pm$0.009~M$_{\odot}$ add 1\% uncertainty on distance in
quadrature., comparable to the 1.3$\pm0.08$M$_{\odot}$ derived using
$^{12}$CO from the IRAM Plateau de Bure interferometer
\citep{Simon2000} and to the 1.23$\pm0.12$~M$_{\odot}$ derived by
\citet{Czekala2019} using ALMA data at 0.6-0.7'' resolution of the
same CO isotopologues in our data. We derive a total disk mass of
0.00078$\pm$0.00003M$_{\odot}$.

\begin{figure}
\begin{center}
\epsscale{0.5}
%\includegraphics[angle=0,scale=0.33]{savefigEmissivity_HIP_73145.pdf}
%\includegraphics[angle=0,scale=0.33]{savefigEmissivity_HIP_84881.pdf}
%\plotone
%\includegraphics[angle=0,scale=0.6]{UZTauE_13CO2-1_data.pdf}
%\includegraphics[angle=0,scale=0.6]{UZTauE_13CO2-1_model.pdf}
%\includegraphics[angle=0,scale=0.6]{UZTauE_13CO2-1_residual.pdf}

\includegraphics[angle=0,scale=0.6]{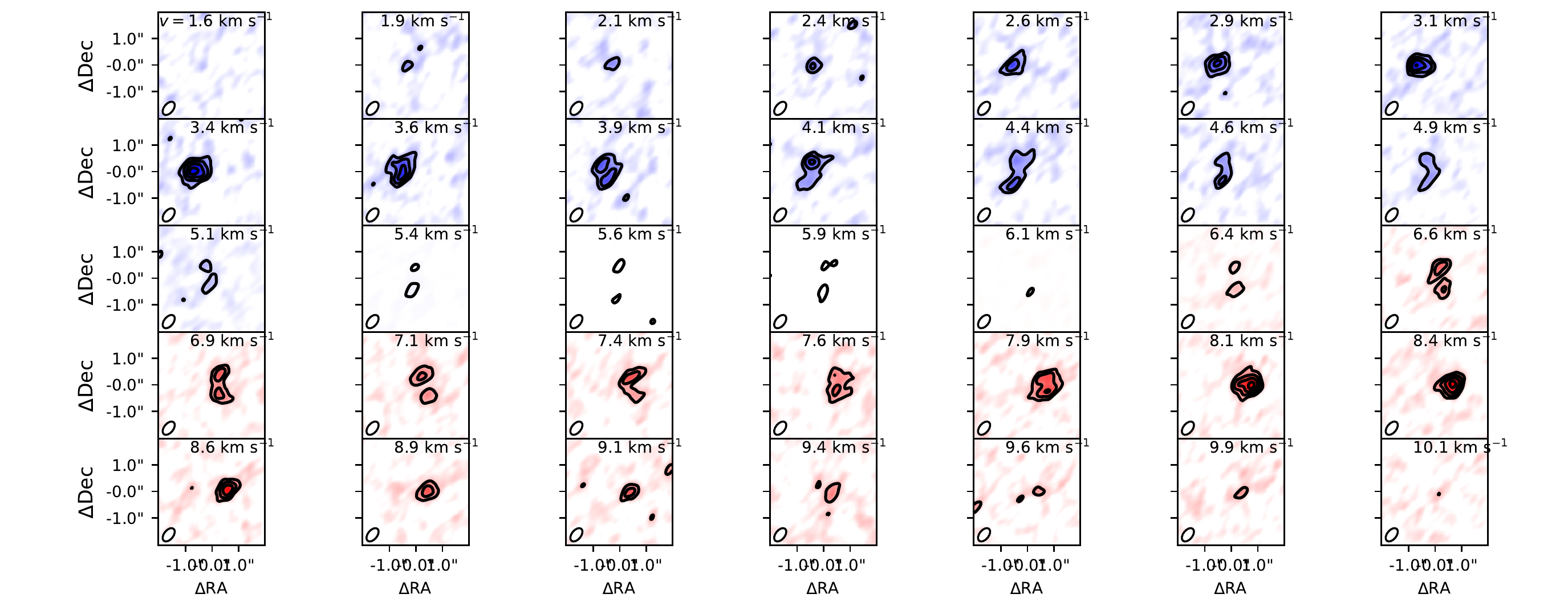}
\includegraphics[angle=0,scale=0.6]{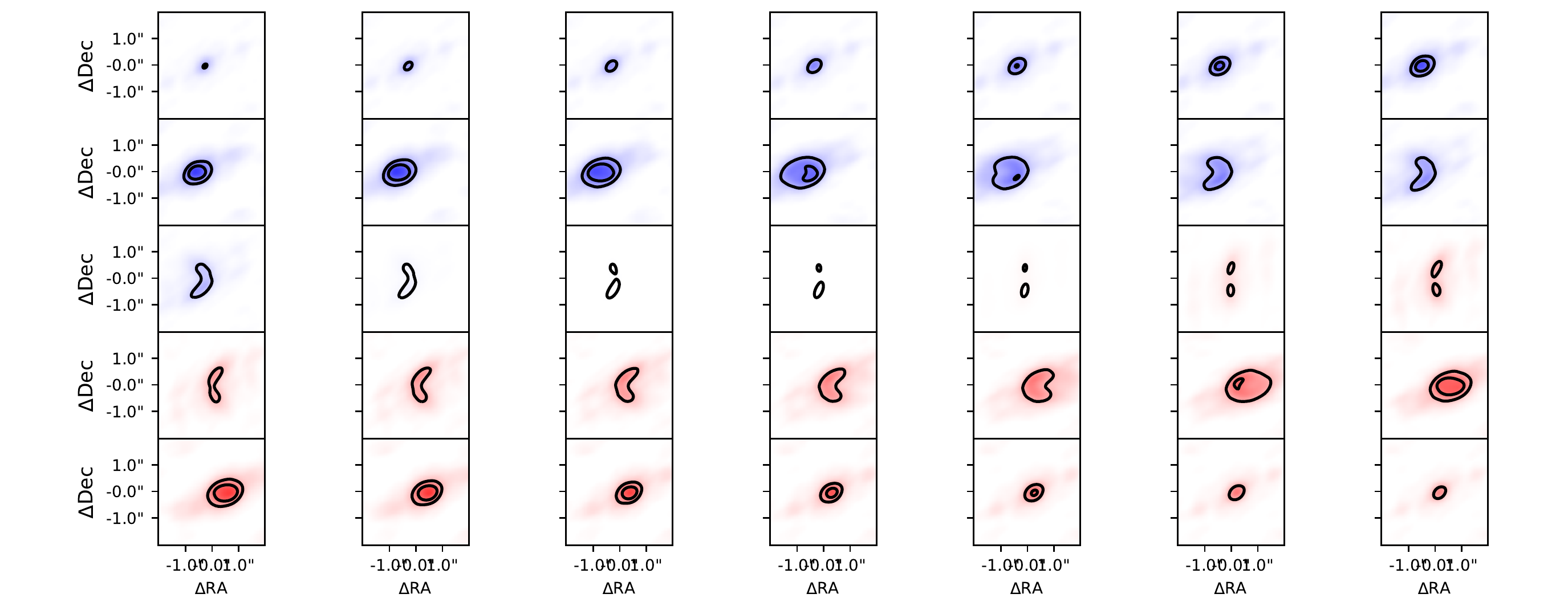}
\includegraphics[angle=0,scale=0.6]{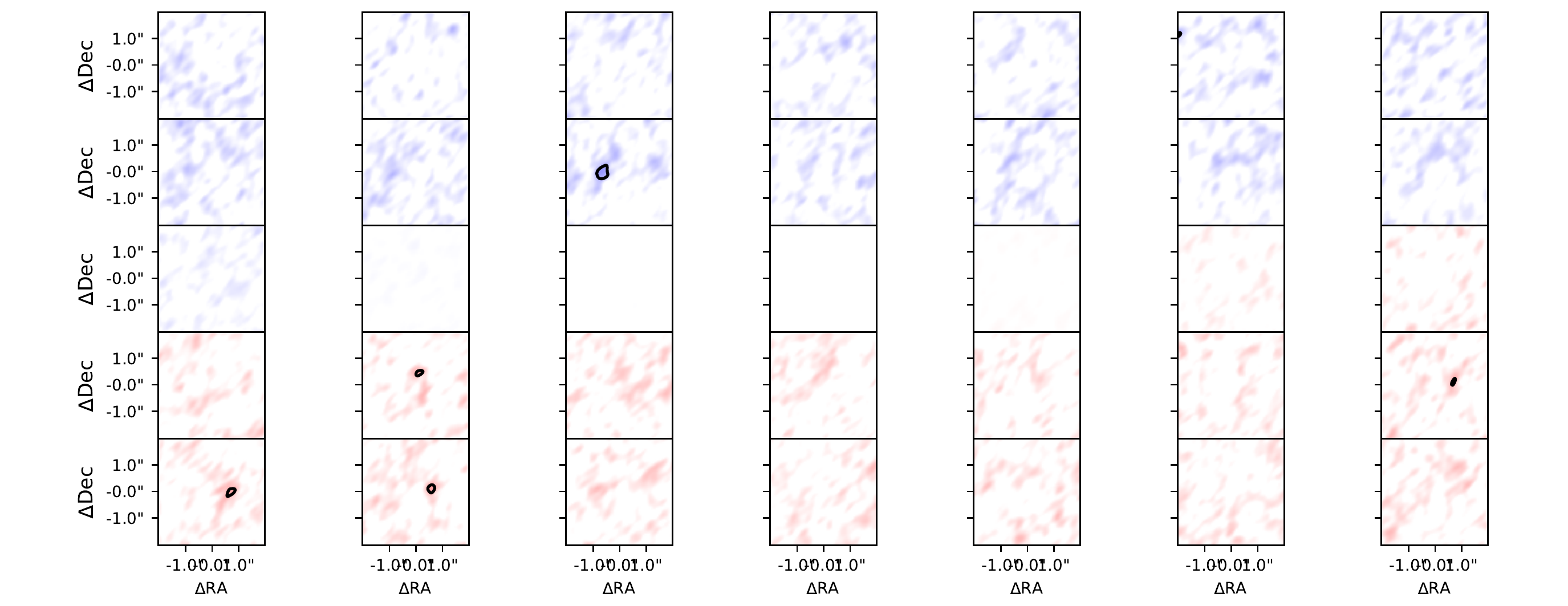}

\caption{$^{13}$CO channel maps for UZ Tau E (top), best-fit disk model (middle) and residuals (bottom). {Contour levels start at 3.5 sigma, with increments of 3 sigma.}}  
\label{fig:mcmc13co}
\end{center}
\end{figure}

\begin{figure}
\begin{center}
\epsscale{0.5}
%\includegraphics[angle=0,scale=0.33]{savefigEmissivity_HIP_73145.pdf}
%\includegraphics[angle=0,scale=0.33]{savefigEmissivity_HIP_84881.pdf}
%\plotone
%\includegraphics[angle=0,scale=0.6]{UZTauE_C18O2-1_data.pdf}
%\includegraphics[angle=0,scale=0.6]{UZTauE_C18O2-1_model.pdf}
%\includegraphics[angle=0,scale=0.6]{UZTauE_C18O2-1_residual.pdf}
\includegraphics[angle=0,scale=0.6]{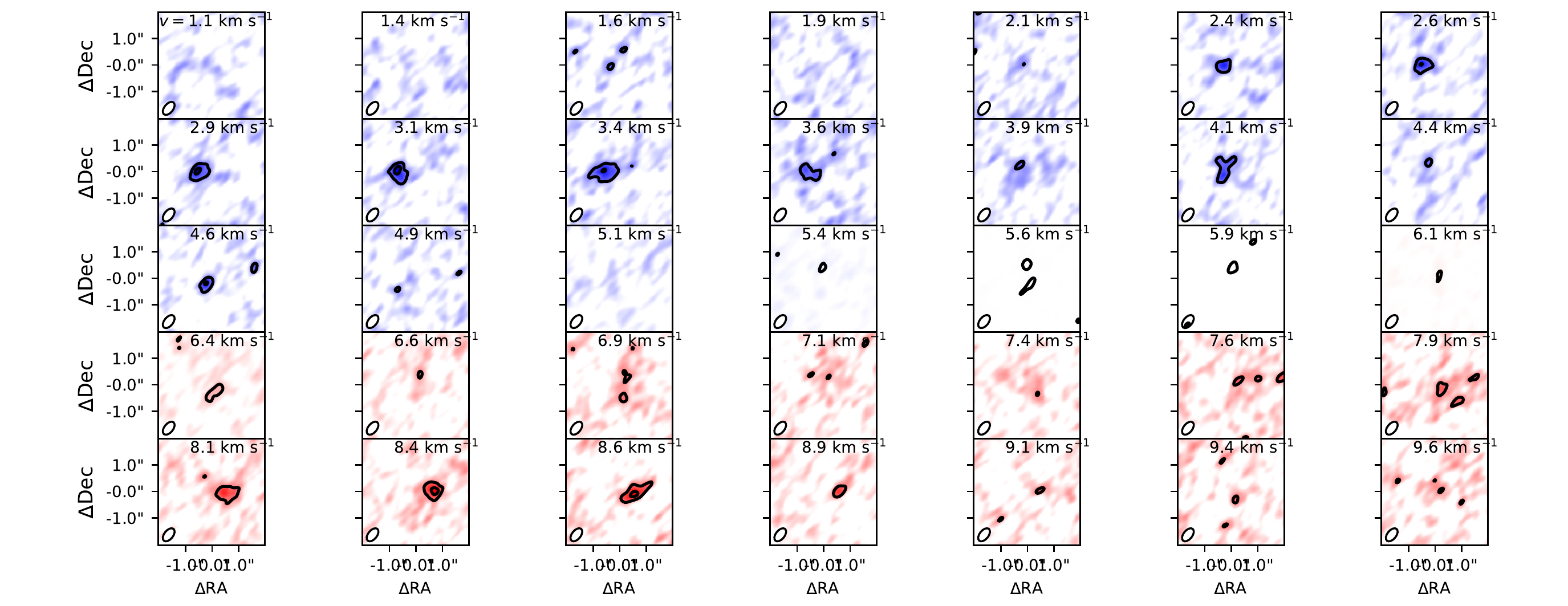}
\includegraphics[angle=0,scale=0.6]{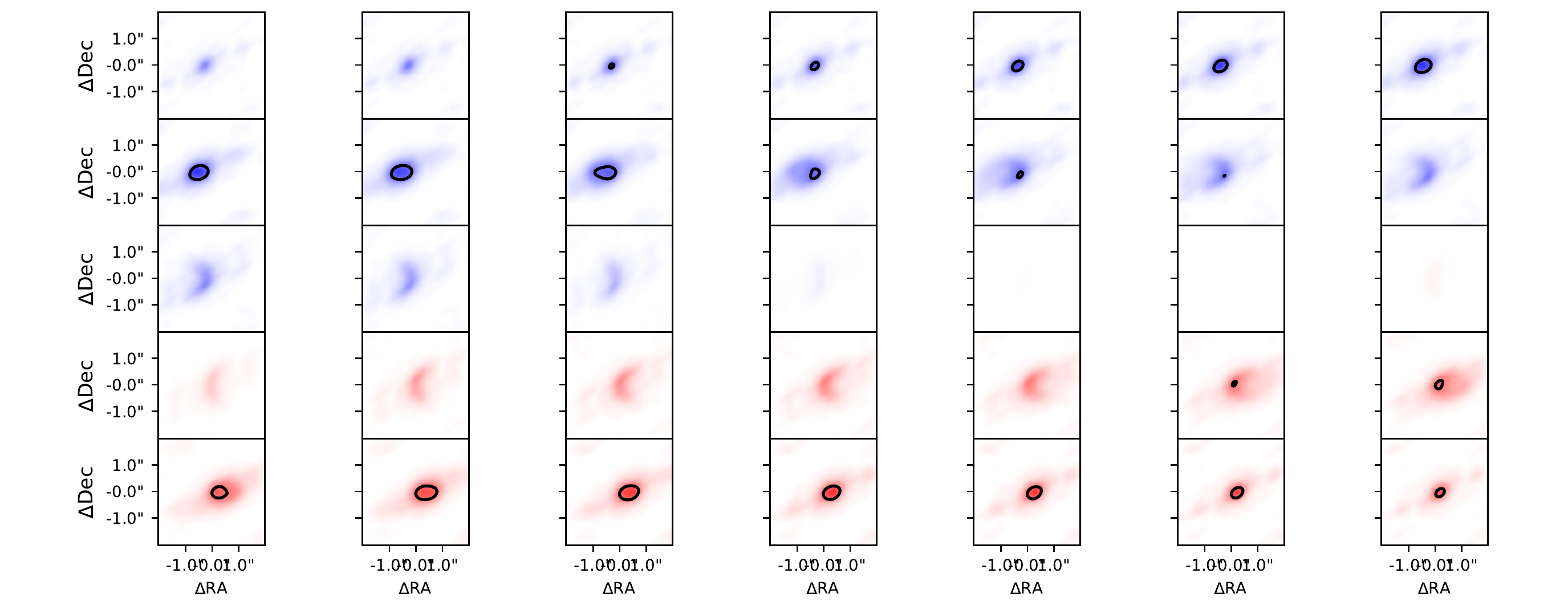}
\includegraphics[angle=0,scale=0.6]{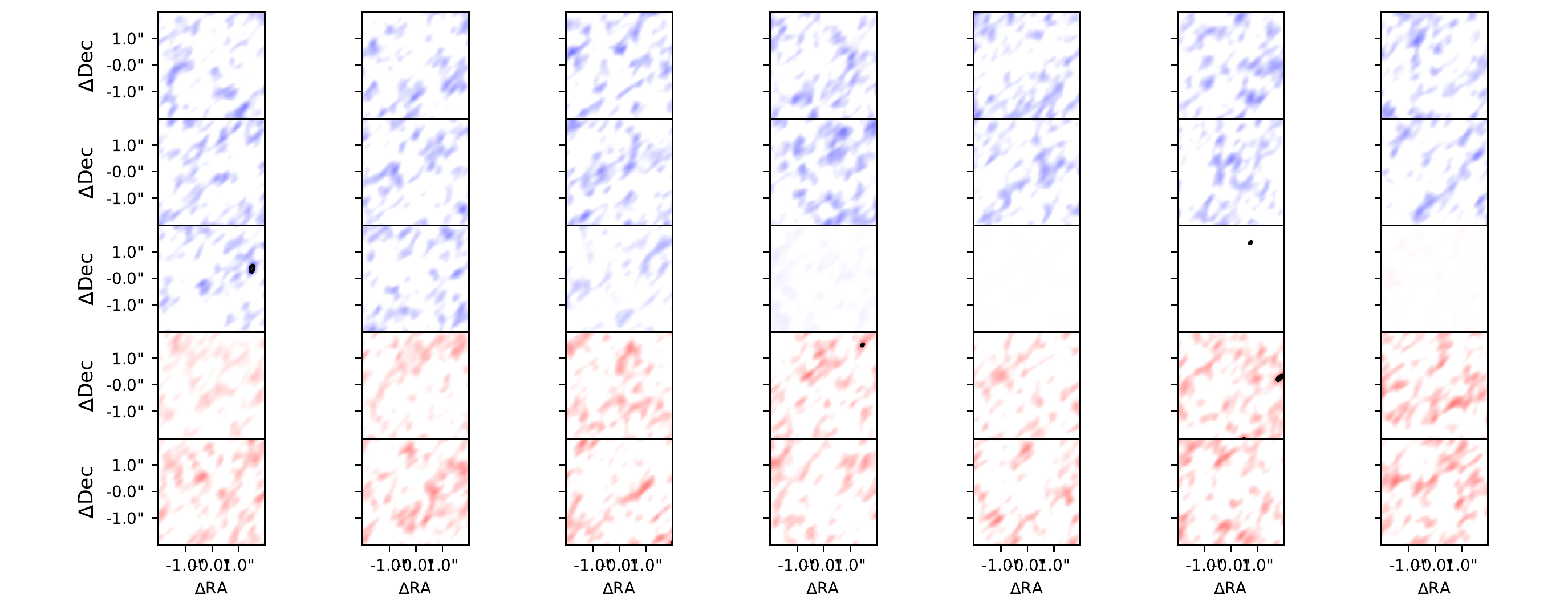}

 \caption{C$^{18}$O channel maps for UZ Tau E (top), best-fit disk model (middle) and residuals (bottom). }  
\label{fig:mcmcc18o}
\end{center}
\end{figure}

\begin{figure}
\begin{center}
\epsscale{0.5}
\includegraphics[angle=0,scale=0.3]{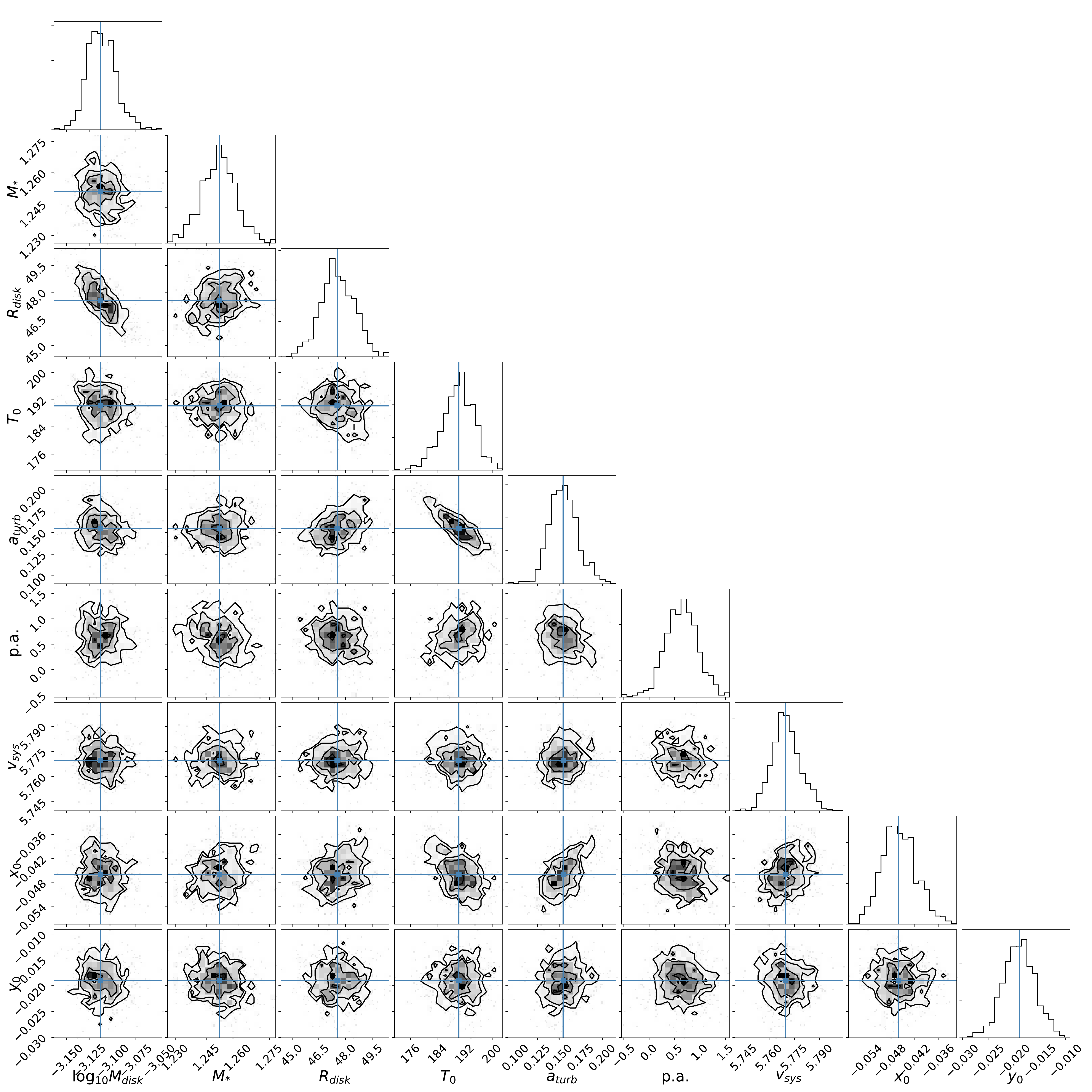}

\caption{Triangle plots of the posterior probability distribution function for the different model parameters.}  
\label{fig:fit}
\end{center}
\end{figure}

\begin{deluxetable*}{lcccccccccccc}
%\tablewidth{700pt}
\tablecaption{Best-fit model parameters for UZ~Tau~E}
%\tabletypesize{\scriptsize}
\tablehead{
\colhead{$\gamma$} & 
\colhead{i} & 
\colhead{M$_{\rm disk}$} &
\colhead{M$_{*}$} &
\colhead{R$_{\rm disk}$} &
\colhead{R$_{\rm in}$} & 
\colhead{T$_{\rm 0}$} &
\colhead{a$_{\rm turb}$} &
\colhead{pa} &
\colhead{q} &
\colhead{v$_{\rm sys}$} &
\colhead{x$_{\rm 0}$} &
\colhead{y$_{\rm 0}$} \\
\colhead{} & 
\colhead{[$^\circ$]} & 
\colhead{[M$_{\odot}$]} &
\colhead{[M$_{\odot}$]} &
\colhead{[au]} &
\colhead{[au]} & 
\colhead{[K]} &
\colhead{[km~s$^{-1}$]} &
\colhead{[$^\circ$]} &
\colhead{} &
\colhead{[km~s$^{-1}$]} &
\colhead{[mas]} &
\colhead{[mas]} \\
}
\startdata
 1.0$^a$ & 56.15$^a$ & 0.00078$\pm$0.00003 & 1.253$\pm$ 0.009&48.1$\pm$1.1  & 1.0$^a$ & 187.3$\pm$4.6& 0.171$\pm$ 0.016&1.1$\pm$0.3 & 0.5$^a$ & 5.77$\pm$ 0.01 &  -0.051 $\pm$0.004 &  -0.013$\pm$0.003 \\
\enddata
\tablecomments{$^a$ Parameter is fixed.}
\label{fit}
\end{deluxetable*}

\section{Channel Maps}\label{co_maps}

\begin{figure*}
\begin{center}
\epsscale{0.5}
\includegraphics[width=0.8\textwidth]{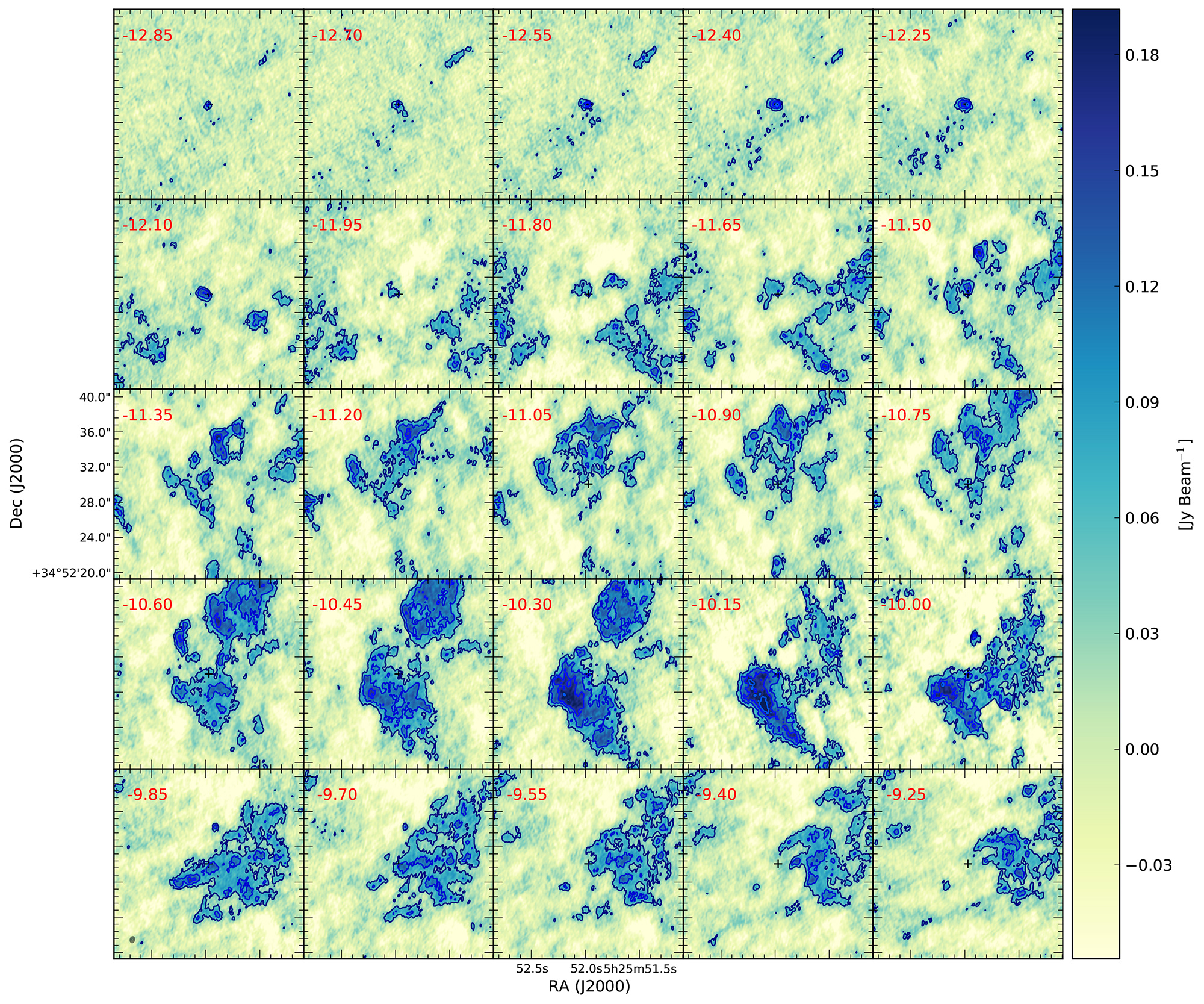}

\caption{ $^{12}$CO channel maps  for V582~Aur. Contour levels start at 3$\sigma$, increasing in steps of 3$\sigma$ ($\sigma$=16~mJy\,beam$^{-1}$). The position of the star is marked with a cross. The beam is represented by a gray ellipse in the bottom left corner of the  bottom left panel.  
}
\label{12cov582}
\end{center}
\end{figure*}

\begin{figure*}
\begin{center}
\epsscale{0.5}
\includegraphics[width=0.8\textwidth]{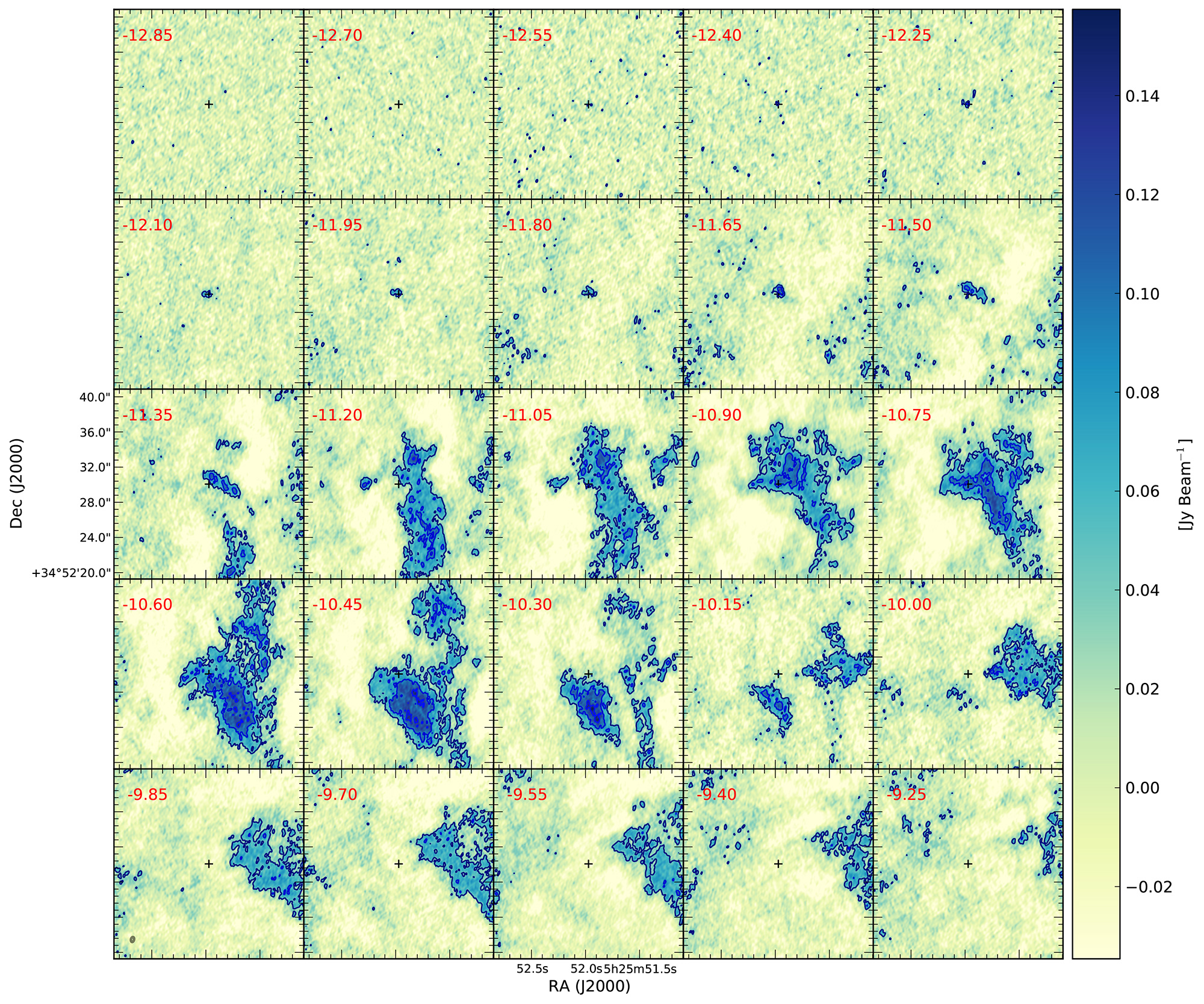}

\caption{ $^{13}$CO channel maps  for V582~Aur. Contour levels start at 3$\sigma$, increasing in steps of 3$\sigma$ ($\sigma$=17~mJy\,beam$^{-1}$). The position of the star is marked with a cross. The beam is represented by a gray ellipse in the bottom left corner of the  bottom left panel.} 
\label{13cov582}
\end{center}
\end{figure*}

\begin{figure*}
\begin{center}
\epsscale{0.5}
\includegraphics[width=0.8\textwidth]{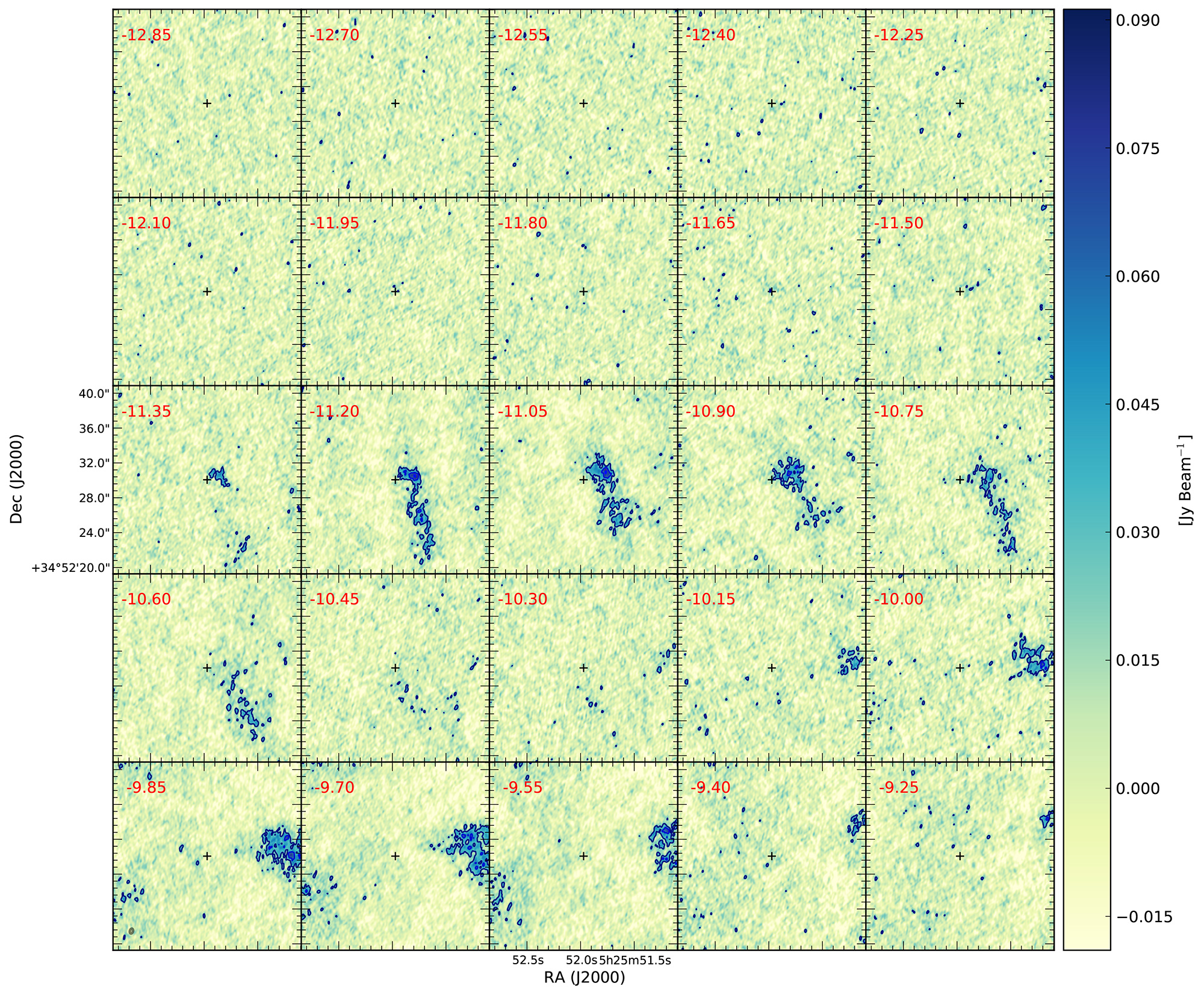}
\caption{ C$^{18}$O channel maps  for V582~Aur. Contour levels start at 3$\sigma$, increasing in steps of 3$\sigma$ ($\sigma$=10~mJy\,beam$^{-1}$). The position of the star is marked with a cross. The beam is represented by a gray ellipse in the bottom left corner of the  bottom left panel.} 
\label{c18ov582}
\end{center}
\end{figure*}

\begin{figure*}
\begin{center}
\epsscale{0.5}
\includegraphics[angle=0,scale=0.3]{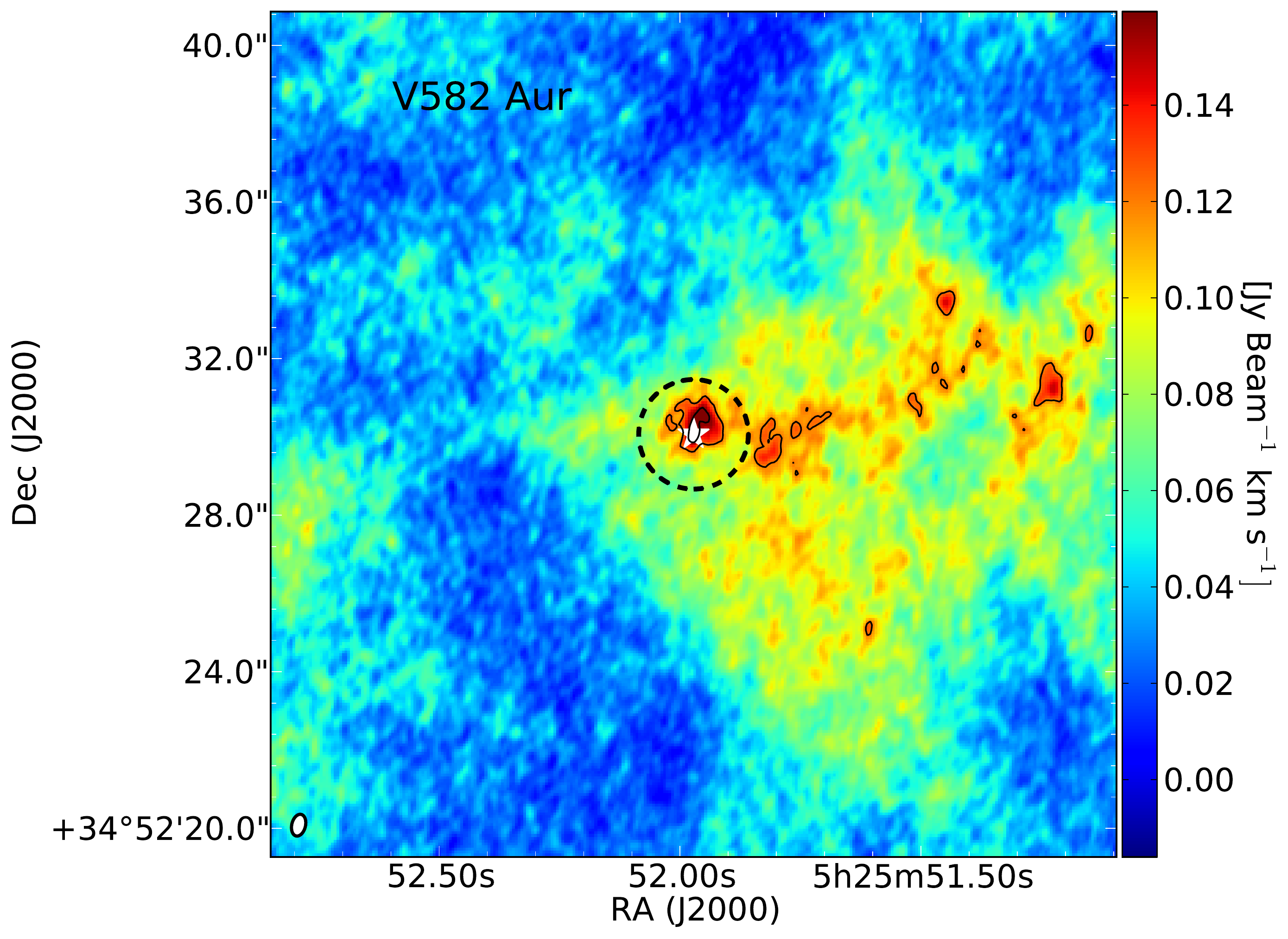}
\includegraphics[angle=0,scale=0.3]{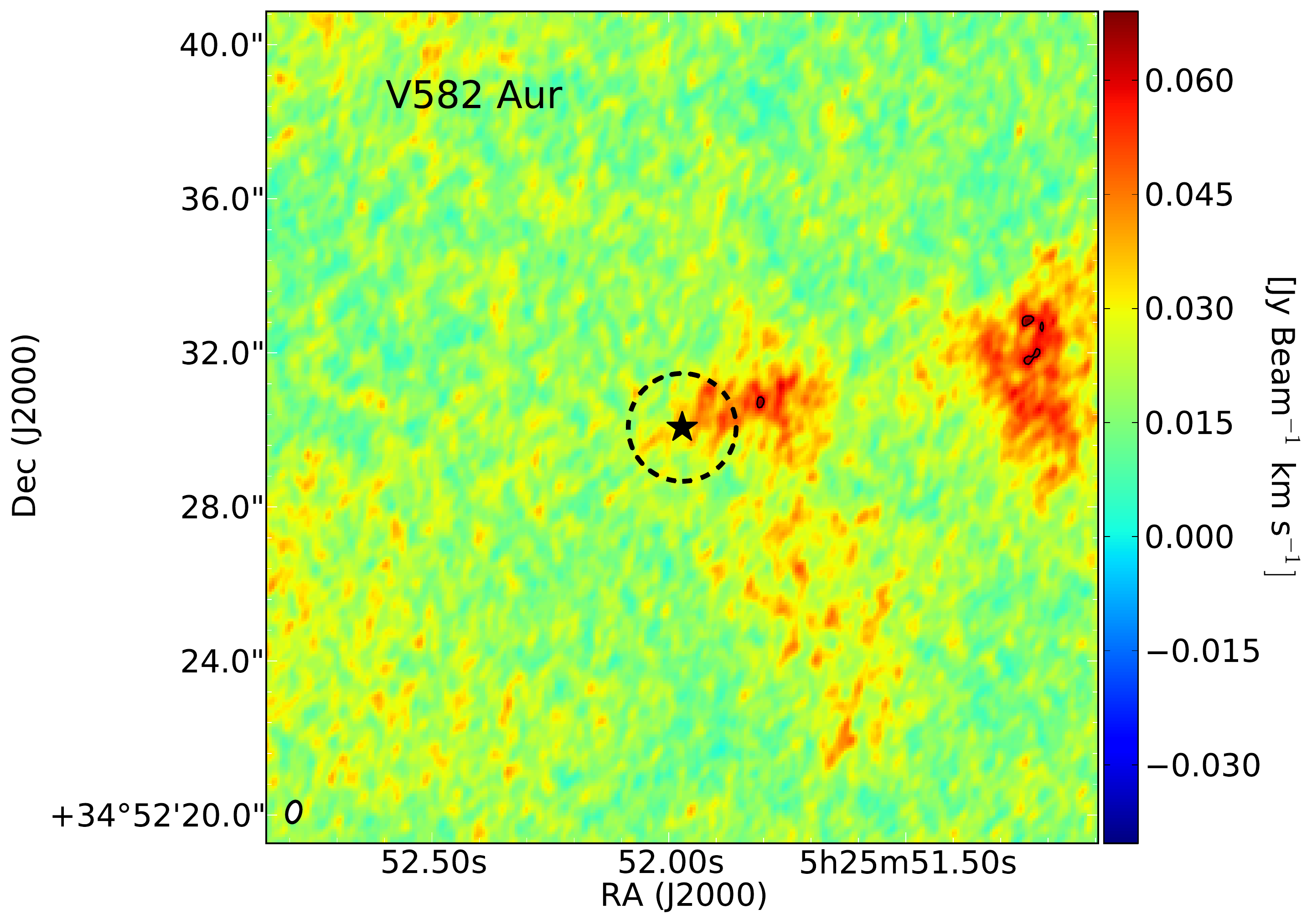}
\caption{  $^{13}$CO(2--1) and C$^{18}$O(2--1) moment~0 maps for V582~Aur (left and right panels, respectively). Contour levels start at 3$\sigma$, increasing in steps of 1$\sigma$. The dashed circles shows the region used to compute the integrated line emissions listed in Table~\ref{cofluxes}. The peak position of the continuum is shown with a star symbol.  The moment~0 maps were computed in the same velocity ranges as Figure~\ref{mom0}.}
\label{fig-mom-v582-iso}
\end{center}

\end{figure*}

Figures~\ref{12cov582}, ~\ref{13cov582} and ~\ref{c18ov582} show the
channel maps for  V582~Aur in $^{12}$CO(2--1), $^{13}$CO(2--1) and C$^{18}$O(2--1)
respectively. Figure~\ref{fig-mom-v582-iso} shows the moment~0 maps for
$^{13}$CO(2--1) and C$^{18}$O(2--1) computed in the same velocity ranges as
Figure~\ref{mom0}.

\begin{thebibliography}{}

  

\bibitem[{\'A}brah{\'a}m et al.(2018)]{abraham2018} {\'A}brah{\'a}m, P., K{\'o}sp{\'a}l, {\'A}., Kun, M., et al.\ 2018, \apj, 853, 28

  
\bibitem[Andrews et al.(2010)]{andrews2010} Andrews, S.~M., Wilner, D.~J., Hughes, A.~M., Qi, C., \& Dullemond, C.~P.\ 2010, \apj, 723, 1241

\bibitem[Andrews et al.(2018)]{andrews2018} Andrews, S.~M., Terrell, M., Tripathi, A., et al.\ 2018, \apj, 865, 157

\bibitem[Andrews(2020)]{andrews2020} Andrews, S.~M.\ 2020, arXiv e-prints, arXiv:2001.05007


\bibitem[Ansdell et al.(2016)]{ansdell2016} Ansdell, M., Williams, J.~P., van der Marel, N., et al.\ 2016, \apj, 828, 46 

  
\bibitem[Ansdell et al.(2018)]{Ansdell2018} Ansdell, M., Williams, J.~P., Trapman, L., et al.\ 2018, \apj, 859, 21

  
\bibitem[Arce \& Sargent(2006)]{arce2006} Arce, H.~G., \& Sargent, A.~I.\ 2006, \apj, 646, 1070 

  
\bibitem[Armitage et al.(2001)]{armitage2001} Armitage, P.~J., Livio, M., \& Pringle, J.~E.\ 2001, \mnras, 324, 705 

\bibitem[Astropy Collaboration et  al.(2013)]{2013A&A...558A..33A} Astropy Collaboration, Robitaille, T.~P., Tollerud, E.~J., et al.\ 2013, \aap, 558, AA33 

  
\bibitem[Audard et al.(2014)]{audard2014} Audard, M., {\'A}brah{\'a}m, P., Dunham, M.~M., et al.\ 2014, Protostars and Planets VI, 387 

%\bibitem[Ballering \& Eisner(2019)]{Ballering2019} Ballering, N.~P., \& Eisner, J.~A.\ 2019, \aj, 157, 144

\bibitem[Bate(2018)]{bate2018} Bate, M.~R.\ 2018, \mnras, 475, 5618

  
\bibitem[Beckwith et al.(1990)]{beckwith1990} Beckwith, S.~V.~W., Sargent, A.~I., Chini, R.~S., \& Guesten, R.\ 1990, \aj, 99, 924 

\bibitem[Bonnell \& Bastien(1992)]{Bonnell1992} Bonnell, I., \& Bastien, P.\ 1992, \apjl, 401, L31

\bibitem[Bosman et al.(2018)]{Bosman2018} Bosman, A.~D., Walsh, C., \& van Dishoeck, E.~F.\ 2018, \aap, 618, A182

\bibitem[Calvet et al.(1993)]{Calvet1993} Calvet, N., Hartmann, L., \& Kenyon, S.~J.\ 1993, The Astrophysical Journal, 402, 623

\bibitem[Chiang \& Goldreich(1997)]{Chiang1997} Chiang, E.~I., \& Goldreich, P.\ 1997, \apj, 490, 368

  
\bibitem[Cieza et al.(2016)]{cieza2016} Cieza, L.~A., Casassus, S., Tobin, J., et al.\ 2016, \nat, 535, 258

\bibitem[Cieza et al.(2018)]{cieza2018} Cieza, L.~A., Ru{\'\i}z-Rodr{\'\i}guez, D., Perez, S., et al.\ 2018, \mnras, 474, 4347.

\bibitem[Cieza et al.(2019)]{cieza2019} Cieza, L.~A., Ru{\'\i}z-Rodr{\'\i}guez, D., Hales, A., et al.\ 2019, \mnras, 482, 698

  
\bibitem[Connelley \& Reipurth(2018)]{connelley2018} Connelley, M.~S., \& Reipurth, B.\ 2018, \apj, 861, 145


\bibitem[Cruz-S{\'a}enz de Miera et al.(2019)]{cruz2019} Cruz-S{\'a}enz de Miera, F., K{\'o}sp{\'a}l, {\'A}., {\'A}brah{\'a}m, P., et al.\ 2019, arXiv e-prints, arXiv:1908.04649
  
\bibitem[Cuello et al.(2019)]{Cuello2019} Cuello, N., Dipierro, G., Mentiplay, D., et al.\ 2019, \mnras, 483, 4114

\bibitem[Czekala et al.(2019)]{Czekala2019} Czekala, I., Chiang, E., Andrews, S.~M., et al.\ 2019, \apj, 883, 22

  

\bibitem[Dewangan et al.(2018)]{Dewangan2018} Dewangan, L.~K., Baug, T., Ojha, D.~K., et al.\ 2018, \apj, 864, 54


\bibitem[Dullemond et al.(2012)]{Dullemond2012} Dullemond, C.~P., Juhasz, A., Pohl, A., et al.\ 2012, RADMC-3D: A multi-purpose radiative transfer tool, ascl:1202.015

\bibitem[Dullemond et al.(2019)]{Dullemond2019} Dullemond, C.~P., K{\"u}ffmeier, M., Goicovic, F., et al.\ 2019, \aap, 628, A20
  
\bibitem[Dzib et al.(2018)]{Dzib2018} Dzib, S.~A., Loinard, L., Ortiz-Le{\'o}n, G.~N., et al.\ 2018, \apj, 867, 151

  
  
\bibitem[Evans et al.(2009)]{evans2009} Evans, N.~J., II, Dunham, M.~M., J{\o}rgensen, J.~K., et al.\ 2009, \apjs, 181, 321-350 

\bibitem[Foreman-Mackey et al.(2013)]{foreman2013} Foreman-Mackey, D.,
  Hogg, D.~W., Lang, D., \& Goodman, J.\ 2013, \pasp, 125, 306
  
\bibitem[Frank et al.(2014)]{Frank2014} Frank, A., Ray, T.~P., Cabrit, S., et al.\ 2014, Protostars and Planets VI, 451

  
\bibitem[Gaia Collaboration et al.(2018)]{gaia2018} Gaia Collaboration, Brown, A.~G.~A., Vallenari, A., et al.\ 2018, \aap, 616, A1 


\bibitem[C. Gonzalez et al.(in prep.)]{gonzalesprep} Gonzalez C., Hales, A.~S, Cieza, L.~A. et al.\ in preparation

  
\bibitem[Grinin et al.(2019)]{Grinin2019} Grinin, V.~P., Semenov, A.~O., Barsunova, O.~Y., et al.\ 2019, Astrophysics, 62, 41

  
\bibitem[Hales et al.(2015)]{hales2015} Hales, A.~S., Corder, S.~A., Dent, W.~R.~D., et al.\ 2015, \apj, 812, 134 
  
\bibitem[Hales et al.(2018)]{hales2018} Hales, A.~S., P{\'e}rez, S., Saito, M., et al.\ 2018, \apj, 859, 111

\bibitem[Hales et al. (in prep.)]{halesprep}  Hales, A.~S, Williams, J.~P., Cieza, L.~A. et al., in preparation

  
\bibitem[Hartmann, \& Kenyon(1985)]{Hartmann1985} Hartmann, L., \& Kenyon, S.~J.\ 1985, \apj, 299, 462

  
\bibitem[Hartmann \& Kenyon(1996)]{hartmann1996} Hartmann, L., \& Kenyon, S.~J.\ 1996, \araa, 34, 207 

\bibitem[Hartmann(2008)]{hartmann2008} Hartmann, L.\ 2008, Accretion Processes in Star Formation, Cambridge, UK: Cambridge University Press, 2008,  

  
\bibitem[Hartmann et al.(2016)]{hartmann2016} Hartmann, L., Herczeg, G., \& Calvet, N.\ 2016, \araa, 54, 135 

\bibitem[Herbig(1966)]{herbig1966} Herbig, G.~H.\ 1966, Vistas in Astronomy, 8, 109 

\bibitem[Herbig, \& Harlan(1971)]{Herbig1971} Herbig, G.~H., \& Harlan, E.~A.\ 1971, Information Bulletin on Variable Stars, 543, 1

\bibitem[Hildebrand(1983)]{Hildebrand1983} Hildebrand, R.~H.\ 1983, \qjras, 24, 267

\bibitem[Huang et al.(2018)]{Huang2018} Huang, J., Andrews, S.~M., Dullemond, C.~P., et al.\ 2018, \apjl, 869, L42

  
\bibitem[Hubbard(2017)]{hubbard2017} Hubbard, A.\ 2017, \mnras, 465, 1910 


\bibitem[Jensen et al.(2007)]{Jensen2007} Jensen, E.~L.~N., Dhital, S., Stassun, K.~G., et al.\ 2007, \aj, 134, 241

   
\bibitem[Jones et al.(1985)]{jones1985} Jones, T.~J., Hyland, A.~R., Harvey, P.~M., et al.\ 1985, \aj, 90, 1191

\bibitem[Johnstone et al.(2018)]{Johnstone2018} Johnstone, D., Herczeg, G.~J., Mairs, S., et al.\ 2018, The Astrophysical Journal, 854, 31

  
\bibitem[Kenyon et al.(1990)]{Kenyon1990} Kenyon, S.~J., Hartmann, L.~W., Strom, K.~M., et al.\ 1990, \aj, 99, 869

\bibitem[K{\'o}sp{\'a}l et al.(2017b)]{kospal2017b} K{\'o}sp{\'a}l, {\'A}., {\'A}brah{\'a}m, P., Csengeri, T., et al.\ 2017, \apj, 843, 45 %V346 Nor

\bibitem[Kospal(2018)]{kospal2018} Kospal, A.\ 2018, Take a Closer Look, 15-19 October, 2018 in ESO-HQ, Garching b. München, Germany. Online at http://www.eso.org/sci/meetings/2018/tcl2018.html, tcl2018, id.87

\bibitem[Krijt et al.(2018)]{Krijt2018} Krijt, S., Schwarz, K.~R., Bergin, E.~A., et al.\ 2018, \apj, 864, 78

  
\bibitem[Kun et al.(2017)]{kun2017} Kun, M., Szegedi-Elek, E., \& Reipurth, B.\ 2017, \mnras, 468, 2325 

\bibitem[Ladd et al.(1998)]{Ladd1998} Ladd, E.~F., Fuller, G.~A., \& Deane, J.~R.\ 1998, \apj, 495, 871

  
\bibitem[Liu et al.(2018)]{liu2018} Liu, H.~B., Dunham, M.~M., Pascucci, I., et al.\ 2018, \aap, 612, A54

\bibitem[Liu et al.(2019)]{liu2019} Liu, H.~B., M{\'e}rand, A., Green, J.~D., et al.\ 2019, \apj, 884, 97

\bibitem[Lodato \& Clarke(2004)]{lodato2004} Lodato, G., \& Clarke, C.~J.\ 2004, \mnras, 353, 841 


\bibitem[Long et al.(2018)]{long2018} Long, F., Pinilla, P., Herczeg, G.~J., et al.\ 2018, \apj, 869, 17


\bibitem[Lorenzetti et al.(2007)]{Lorenzetti2007} Lorenzetti, D., Giannini, T., Larionov, V.~M., et al.\ 2007, \apj, 665, 1182


\bibitem[Lynden-Bell, \& Pringle(1974)]{lyndelbell+1974} Lynden-Bell, D., \& Pringle, J.~E.\ 1974, MNRAS, 168, 603

\bibitem[Manara et al.(2019)]{manara2019} Manara, C.~F., Tazzari, M., Long, F., et al.\ 2019, \aap, 628, A95

\bibitem[Marino et al.(2017)]{Marino2017} Marino, S., Wyatt, M.~C., Pani{\'c}, O., et al.\ 2017, \mnras, 465, 2595


\bibitem[Moody, \& Stahler(2017)]{Moody2017} Moody, M.~S.~L., \& Stahler, S.~W.\ 2017, \aap, 600, A133
  
\bibitem[Mathieu et al.(1996)]{mathieu1996} Mathieu, R.~D., Martin, E.~L., \& Magazzu, A.\ 1996, American Astronomical Society Meeting Abstracts \#188 188, 60.05


\bibitem[Mattila et al.(1989)]{mattila1989} Mattila, K., Liljestr{\"o}m, T., \& Toriseva, M.\ 1989, European Southern Observatory Conference and Workshop Proceedings, 153

    
\bibitem[McMullin et al.(2007)]{2007ASPC..376..127M} McMullin, J.~P., Waters, B., Schiebel, D., Young, W., \& Golap, K.\ 2007, Astronomical Data Analysis Software and Systems XVI, 376, 127

\bibitem[Miotello et al.(2017)]{Miotello2017} Miotello, A., van Dishoeck, E.~F., Williams, J.~P., et al.\ 2017, \aap, 599, A113
  
\bibitem[Mottram et al.(2017)]{Mottram2017} Mottram, J.~C., van Dishoeck, E.~F., Kristensen, L.~E., et al.\ 2017, \aap, 600, A99

\bibitem[Perez et al.(2015)]{perez2015} Perez, S., Dunhill, A., Casassus, S., et al.\ 2015, \apjl, 811, L5



\bibitem[P{\'e}rez et al.(2018)]{perez2018} P{\'e}rez, S., Casassus, S., \& Ben{\'\i}tez-Llambay, P.\ 2018, \mnras, 480, L12

\bibitem[P{\'e}rez et al.(2020)]{perez2020} P{\'e}rez, S., Hales, A., Liu, H.~B., et al.\ 2020, \apj, 889, 59
  
\bibitem[Persi et al.(2007)]{persi2007} Persi, P., Tapia, M., G{\`o}mez, M., et al.\ 2007, \aj, 133, 1690

\bibitem[Pinte et al.(2016)]{Pinte2016} Pinte, C., Dent, W.~R.~F., M{\'e}nard, F., et al.\ 2016, \apj, 816, 25

  
\bibitem[Prato et al.(2002)]{Prato2002} Prato, L., Simon, M., Mazeh, T., et al.\ 2002, \apjl, 579, L99

  
\bibitem[Principe et al.(2018)]{principe2018} Principe, D.~A., Cieza, L., Hales, A., et al.\ 2018, \mnras, 473, 879 


\bibitem[Reipurth et al.(2012)]{Reipurth2012} Reipurth, B., Aspin, C., \& Herbig, G.~H.\ 2012, \apj, 748, L5

\bibitem[Reipurth, \& Aspin(2010)]{Reipurth2010} Reipurth, B., \& Aspin, C.\ 2010, Evolution of Cosmic Objects Through Their Physical Activity, 19


\bibitem[Rosotti et al.(2019)]{Rosotti2019} Rosotti, G.~P., Booth, R.~A., Tazzari, M., et al.\ 2019, \mnras, 486, L63


  
\bibitem[Ru{\'{\i}}z-Rodr{\'{\i}}guez et al.(2017a)]{ruiz2017a} Ru{\'{\i}}z-Rodr{\'{\i}}guez, D., Cieza, L.~A., Williams, J.~P., et al.\ 2017, \mnras, 466, 3519 

\bibitem[Ru{\'\i}z-Rodr{\'\i}guez et al.(2017b)]{ruiz2017b} Ru{\'\i}z-Rodr{\'\i}guez, D., Cieza, L.~A., Williams, J.~P., et al.\ 2017, \mnras, 468, 3266.

\bibitem[Safron et al.(2015)]{Safron2015} Safron, E.~J., Fischer, W.~J., Megeath, S.~T., et al.\ 2015, \apjl, 800, L5

  
\bibitem[Samus(2009)]{Samus2009} Samus, N.\ 2009, Central Bureau Electronic Telegrams 1896, 1


\bibitem[Schwarz et al.(2018)]{Schwarz2018} Schwarz, K.~R., Bergin, E.~A., Cleeves, L.~I., et al.\ 2018, \apj, 856, 85


\bibitem[Segura-Cox et al.(2018)]{Segura-Cox2018} Segura-Cox, D.~M., Looney, L.~W., Tobin, J.~J., et al.\ 2018, \apj, 866, 161

\bibitem[Tobin et al.(2020)]{tobin2020} Tobin, J.~J., Sheehan, P., Megeath, S.~T., et al.\ 2020, arXiv e-prints, arXiv:2001.04468

  
\bibitem[Semkov et al.(2013)]{Semkov2013} Semkov, E.~H., Peneva, S.~P., Munari, U., et al.\ 2013, \aap, 556, A60

\bibitem[Sheehan et al.(2019)]{Sheehan2019} Sheehan, P.~D., Wu, Y.-L., Eisner, J.~A., et al.\ 2019, The Astrophysical Journal, 874, 136

\bibitem[Simon et al.(2000)]{Simon2000} Simon, M., Dutrey, A., \& Guilloteau, S.\ 2000, \apj, 545, 1034

\bibitem[Sipos et al.(2009)]{sipos2009} Sipos, N., {\'A}brah{\'a}m, P., Acosta-Pulido, J., et al.\ 2009, \aap, 507, 881 
  

\bibitem[Takami et al.(2018)]{takami2018} Takami, M., Fu, G., Liu, H.~B., et al.\ 2018, The Astrophysical Journal, 864, 20

\bibitem[Takami et al.(2019)]{takami2019} Takami, M., Chen, T.-S., Liu, H.~B., et al.\ 2019, \apj, 884, 146


    
\bibitem[Tazzari et al.(2017)]{tazzari2017} Tazzari, M., Beaujean, F., \& Testi, L.\ 2017, galario: Gpu Accelerated Library for Analyzing Radio Interferometer Observations, ascl:1710.022

    
\bibitem[Tazzari et al.(2018)]{tazzari2018} Tazzari, M., Beaujean, F., \& Testi, L.\ 2018, Monthly Notices of the Royal Astronomical Society, 476, 4527

    
\bibitem[Thommes et al.(2011)]{Thommes2011} Thommes, J., Reipurth, B., Aspin, C., et al.\ 2011, Central Bureau Electronic Telegrams 2795, 1

\bibitem[Trapman et al.(2019)]{Trapman2019} Trapman, L., Facchini, S., Hogerheijde, M.~R., et al.\ 2019, \aap, 629, A79

\bibitem[Tripathi et al.(2017)]{tripathi2017} Tripathi, A., Andrews, S.~M., Birnstiel, T., et al.\ 2017, \apj, 845, 44

  
\bibitem[Tripathi et al.(2018)]{tripathi2018} Tripathi, A., Andrews, S.~M., Birnstiel, T., et al.\ 2018, \apj, 861, 64

  
\bibitem[Vorobyov \& Basu(2015)]{vorobyov2015} Vorobyov, E.~I., \& Basu, S.\ 2015, \apj, 805, 115 

\bibitem[Williams \& Cieza(2011)]{williams2011} Williams, J.~P., \& Cieza, L.~A.\ 2011, \araa, 49, 67

\bibitem[Williams \& Best(2014)]{williams2014} Williams, J.~P., \& Best, W.~M.~J.\ 2014, \apj, 788, 59 

\bibitem[Wilson \& Rood(1994)]{Wilson1994} Wilson, T.~L., \& Rood, R.\ 1994, \araa, 32, 191

  
\bibitem[Zhu et al.(2009a)]{zhu2009a} Zhu, Z., Espaillat, C., Hinkle, K., et al.\ 2009, \apjl, 694, L64

\bibitem[Zhu et al.(2009b)]{zhu2009b} Zhu, Z., Hartmann, L., Gammie, C., et al.\ 2009, \apj, 701, 620

  
\bibitem[Zsidi et al.(2019)]{Zsidi2019} Zsidi, G., {\'A}brah{\'a}m, P., Acosta-Pulido, J.~A., et al.\ 2019, \apj, 873, 130

  
\bibitem[Zurlo et al.(2017)]{zurlo2017} Zurlo, A., Cieza, L.~A., Williams, J.~P., et al.\ 2017, \mnras, 465, 834

\bibitem[Zurlo et al.(in prep.)]{zurloprep} Zurlo, A., Cieza, L.~A., Williams, J.~P., et al.\ in preparation

  
\end{thebibliography}
\end{document}